\documentclass[preprint,showpacs,preprintnumbers,amsmath,amssymb]{revtex4}

\usepackage{graphicx}
\usepackage{dcolumn}
\usepackage{bm}

\newcommand{\mus}{\mbox{$\mu$s}}

\newcommand{\us}{\mbox{$\mu$s}}

\newcommand{\Hn}{\mbox{$^{1}$H}}
\newcommand{\Og}{\mbox{$^{16}$O}}
\newcommand{\C}{\mbox{$^{12}$C}}

\newcommand{\Cd}{\mbox{$^{13}$C}}

\newcommand{\N}{\mbox{$^{12}$N}}

\newcommand{\Ngs}{\mbox{$^{12}$N$_{\rm g.s.}$}}

\newcommand{\B}{\mbox{$^{12}$B}}

\newcommand{\Bu}{\mbox{$^{11}$B}}
\newcommand{\Bgs}{\mbox{$^{12}$B$_{\rm g.s.}$}}
\newcommand{\Fe}{\mbox{$^{56}$Fe}}
\newcommand{\Co}{\mbox{$^{56}$Co}}

\newcommand{\numu}{\mbox{$\nu_{\mu}$}}
\newcommand{\numub}{\mbox{$\bar{\nu}_{\mu}$}}
\newcommand{\nue}{\mbox{$\nu_{e}$}}

\newcommand{\nueb}{\mbox{$\bar{\nu}_{e}$}}
\newcommand{\nutau}{\mbox{$\nu_{\tau}$}}

\newcommand{\nux}{\mbox{$\nu_{x}$}}

\newcommand{\ep}{\mbox{e$^{+}$}}

\newcommand{\el}{\mbox{e$^{-}$}}
\newcommand{\pos}{\mbox{e$^{+}$}}

\newcommand{\mum}{\mbox{$\mu^{-}$}}
\newcommand{\mup}{\mbox{$\mu^{+}$}}
\newcommand{\mupm}{\mbox{$\mu^{\pm}$}}
\newcommand{\pim}{\mbox{$\pi^{-}$}}
\newcommand{\pip}{\mbox{$\pi^{+}$}}

\newcommand{\mupdecay}{\mbox{\mup\ $\rightarrow\:$ \pos $\!$ + \nue\ + \numub}}
\newcommand{\mupdeb}{\mbox{\mup\ $\rightarrow\:$ \pos $\!$ + \nueb\ + \numu}}

\newcommand{\Bdecay}{\mbox{\Bgs\ $\rightarrow\:$ \C\ + \el\ + \nueb}}

\newcommand{\mucap}{\mbox{\mum\ + \C$\ \rightarrow$ \B\ + \numu}}

\newcommand{\mucapxn}{\mbox{\mum\ + \C$\ \rightarrow$ $^{12-x}$B + x$\cdot $n + \numu}}

\newcommand{\Fen}{\mbox{ \Fe\,(\,\mum\,n\,)\,$^{55}$Mn }}

\newcommand{\nuebanp}{\mbox{ p\,(\,\nueb\,,\,$e^+$\,)\,n }}
\newcommand{\nuebanc}{\mbox{ \C\,(\,\nueb\,,\,$e^+$\,n)\,$^{11}$B }}

\newcommand{\CC}{\mbox{\C\,(\,\nue\,,\,\el\,)\,\N }}

\newcommand{\CCprot}{\mbox{p\,(\,\nueb\,,\,\ep\,)\,n }}
\newcommand{\excl}{\mbox{\C\,(\,\nue\,,\,\el\,)\,\N$_{\rm g.s.}$}}
\newcommand{\CCexc}{\mbox{\C\,(\,\nue\,,\,\el\,)\,\N$^{*}$}}

\newcommand{\CCFe}{\mbox{\Fe\,(\,\nue\,,\,\el\,)\,\Co }}
\newcommand{\CCFen}{\mbox{\Fe\,(\,\nue\,,\,\el\,n)\,\Co }}
\newcommand{\NC}{\mbox{\C\,(\,$\nu$\,,\,$\nu^\prime$\,)\,\C$^{*}$}}

\newcommand{\isov}{\mbox{\C\,(\,$\nu$\,,\,$\nu^\prime$\,)\,\C$^{*}$
(\,1$^+$\,,\,1\,;\,15.1 MeV\,) }}

\newcommand{\Nzg}{\mbox{$^{12}$N$_{\rm g.s.}$}}
\newcommand{\nuex}{\mbox{\nue $\rightarrow\,$\nux }}

\newcommand{\numunutau}{\mbox{\numu $\rightarrow\,$\nutau }}

\newcommand{\numunue}{\mbox{\numu $\rightarrow\,$\nue }}
\newcommand{\numubnueb}{\mbox{\numub $\rightarrow\,$\nueb }}

\newcommand{\numux}{\mbox{\numu $\rightarrow\,$\nux }}

\newcommand{\nuenueb}{\mbox{\nue $\rightarrow\,$\nueb }}

\newcommand{\NCL}{\mbox{$90\%\,CL$ }}

\newcommand{\Dm}{\mbox{$\Delta m^2$}}
\newcommand{\sit}{\mbox{$\sin^2(2\Theta )$}}
\newcommand{\eV}{\mbox{eV$^2$}}
\newcommand{\Gdng}{\mbox{Gd\,(\,n,$\gamma$\,)}}

\newcommand{\pnd}{\mbox{p\,(\,n,$\gamma$\,)\,d}}

\renewcommand{\[}{\begin{equation}}
\renewcommand{\]}{\end{equation}}
\newcommand{\CBn}{\mbox{\C\,(\,\nueb\,,\,\ep\,n\,)\,\Bu }}

\begin{document}


\title{Upper limits for neutrino oscillations \numubnueb\  \\
from muon decay at rest}


\author{B.~Armbruster,$^1$ I.M.~Blair,$^2$ B.A.~Bodmann,$^3$ N.E.~Booth,$^4$
G.~Drexlin,$^1$ J.A.~Edgington,$^2$\\
 C.~Eichner,$^5$
K.~Eitel,$^1$ E.~Finckh,$^3$ H.~Gemmeke,$^6$ J.~H\"o\ss l,$^3$ T.~Jannakos,$^1$\\
P.~J\"unger,$^3$ M.~Kleifges,$^6$ J.~Kleinfeller,$^1$
W.~Kretschmer,$^3$ R.~Maschuw,$^{1,5}$ C.~Oehler,$^1$
\\P.~Plischke,$^1$ J. Reichenbacher,$^1$ C.~Ruf,$^5$
M.~Steidl,$^1$ J.~Wolf,$^7$ B.~Zeitnitz$^{1,7}$\\
(KARMEN Collaboration)}

 \affiliation {$^1$ Institut f\"ur Kernphysik,
Forschungszentrum Karlsruhe,
  Postfach 3640, D-76021 Karlsruhe, Germany \\
  $^2$ Physics Department, Queen Mary, University London,
  Mile End Road, London E1 4NS, United Kingdom \\
  $^3$ Physikalisches Institut, Universit\"at Erlangen-N\"urnberg,
  Erwin Rommel Strasse 1, D-91058 Erlangen, Germany \\
  $^4$ Department of Physics, University of Oxford,
  Keble Road, Oxford OX1 3RH, United Kingdom \\
  $^5$\it Institut f\"ur Strahlen- und Kernphysik, Universit\"at Bonn,
  Nu\ss allee 14-16, D-53115 Bonn, Germany \\
  $^6$ Institut f\"ur Prozessdatenverarbeitung und Elektronik,
Forschungszentrum Karlsruhe,
  Postfach 3640, D-76021 Karlsruhe, Germany \\
  $^7$ Institut f\"ur experimentelle Kernphysik, Universit\"at Karlsruhe,
  Gaedestr.1, D-76128 Karlsruhe, Germany
}%


\begin{abstract}
The KARMEN experiment at the spallation neutron source ISIS  used
\numub \ from \mup--decay at rest in the search for neutrino
oscillations \numubnueb\ in the appearance mode, with \nuebanp\ as
detection reaction of \nueb . In total, 15 candidates fulfill all
conditions for the \nueb\ signature, in agreement with the
background expectation of 15.8$\pm$0.5 events, yielding no
indication for oscillations. A single event based likelihood
analysis leads to upper limits on the oscillation parameters:
$\sit<1.7\cdot10^{-3}$ for \Dm $\ge$100\,\eV and \Dm$<$0.055\,\eV\
for \sit=1 at 90\% confidence. Thus, KARMEN does not confirm the
LSND experiment and restricts significantly its favored parameter
region for \numubnueb .
\end{abstract}

\maketitle

\section{Introduction}
The study of neutrino masses and mixing originating from
extensions of the Standard Model (SM) is one of the most
interesting issues in particle physics which has also considerable
impact on astrophysical and cosmological problems. For example,
neutrino masses in the range of a few eV would mean a significant
contribution to the matter content in the universe. In addition,
understanding the mass and mixing scheme of neutrinos is a very
promising tool to improve our knowledge on mass generating
mechanisms for all elementary particles.

A very sensitive way of probing neutrino masses and the mixing
between different neutrino flavors is the search for neutrino
oscillations. The experimental progress in this field during the
recent years has been remarkable, yielding strong evidence for
neutrino oscillations from investigations of solar and atmospheric
neutrinos. The long--standing problem of the solar $\nu$--deficit,
observed by different experiments \cite{solrev} including the
latest results from the Sudbury Neutrino Observatory (SNO)
\cite{SNO00}, is consistently explained as the transition of \nue\
into other active neutrino flavors \cite{SNO01},\cite{Bah01}. In
addition, the atmospheric neutrino anomaly gives evidence for
neutrino oscillations, namely for \numux \ disappearance
oscillations \cite{atmrev}. Due to the precision measurements of
the Super--Kamiokande experiment, the oscillation channel
\numunutau \ is strongly favored \cite{atmster}.

Despite the convincing results from solar and atmospheric
$\nu$--oscillation experiments, all indications for oscillations
are obtained by searches in the {\it disappearance} mode. Up to
now, there is only one piece of evidence for $\nu$--oscillations
in the {\it appearance} mode: the LSND (Liquid Scintillator
Neutrino Detector) experiment \cite{LSNDdet} at the Los Alamos
Neutron Science Center (LANSCE) reported 1995 initial results of
the search for \numubnueb \ oscillations with \numub\ produced in
\mup\ decays at rest \cite{lsnd95}. Supported by a positive signal
in the \numunue\ channel \cite{lsnd97}, updates with increased
statistics \cite{lsnd96,lsnd96lett}  underlined the evidence of an
observed \nueb\ excess but also reduced the original signal
strength. The \nueb\ signal is explained as originating from
\numubnueb\ oscillations with an oscillation probability
$P=(0.264\pm0.067\pm0.045)\% $ \cite{lsndfinal}.

Due to the sensitivity region of LSND, these findings suggest
rather high mass differences of $\Dm>0.1$ \eV, which would imply
significant contributions of neutrinos to the cosmological problem
of dark matter. Due to the high \Dm\ scale it is not possible to
accommodate all three evidences (solar, atmospheric, LSND) with
their distinct regions of \Dm\ within the framework of the SM with
its three neutrino flavors, extended by allowing for non--zero
neutrino masses. Proposed solutions to this problem include e.g.
the incorporation of a sterile neutrino state
\cite{suematsu,bereshiani,foot}, supersymmetry \cite{Fae01} or CPT
violation \cite{CPT}. These deep impacts on particle and
astrophysical aspects therefore require a thorough and independent
test of the \numubnueb\ evidence of LSND.

This paper describes the search for \numubnueb\ oscillations by
the KARMEN ({\bf KA}rlsruhe {\bf R}utherford {\bf M}edium {\bf
E}nergy {\bf N}eutrino) experiment, which was located at the
highly pulsed spallation neutron source ISIS of the Rutherford
Laboratory (UK). The results presented here are based on the final
data set recorded with the full experimental setup of KARMEN~2
from February 1997 until March 2001.

The KARMEN experiment took data, in a different experimental
configuration (KARMEN~1), since 1990. In this first period, the
data analysis was focused on the investigation of
neutrino--nucleus interactions \cite{K_cc,Kncnue,Kncnumu}, but
also on the search for the oscillation channels \numunue\
\cite{Zeitnitz} and \nuex\ \cite{Knuenutau}. Other searches of non
standard model physics such as new particles in pion decay
\cite{Kpeak}, lepton flavor violating pion and muon decays
\cite{KEPhD} or non \mbox{V--A} contributions to the muon decay
\mupdecay\ \cite{K_omega} were also performed. Here, we report on
the most sensitive channel, the search for \numubnueb\
oscillations.

The paper is organized as follows: Section~\ref{sec_isis}
describes the neutrino source ISIS and the KARMEN detector, after
which, in section~\ref{sec_signal}, the \numubnueb\ oscillation
signature is presented. Section~\ref{subsec_req} defines some
general event requirements for the identification of
\nueb--induced events in the data analysis. We discuss the
background in section~\ref{sec_bg}. The final event sample
together with the final data cuts and background expectations is
given in section~\ref{sec_eva}. The data analysis is described in
detail in section~\ref{subsec_ana} together with the presentation
of the final \numubnueb\ results. A detailed discussion of the
results with respect to the LSND evidence and the negative results
from other experiments follows in section \ref{subsec:Comp}.
\section{Experimental Configuration \label{sec_isis}}
\subsection{The neutrino source ISIS}
\begin{figure*}

\begin{center}
\includegraphics{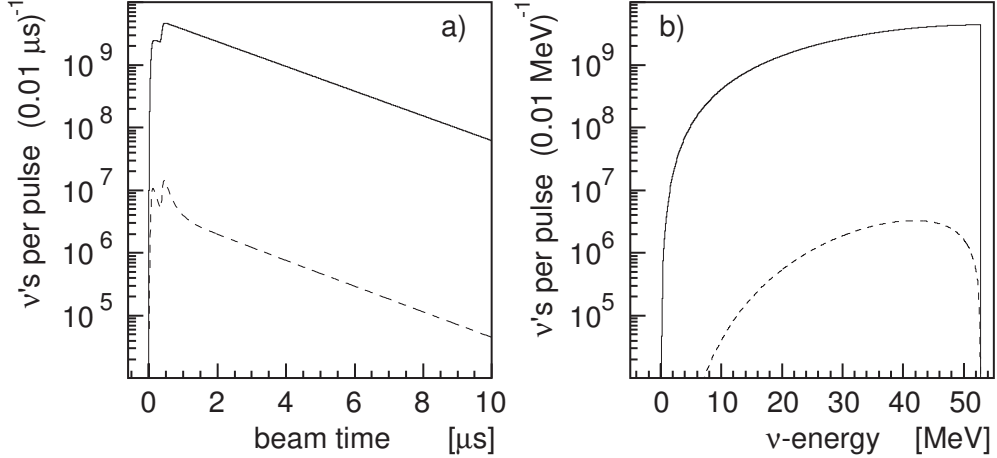}
\caption{ (a) Time  and (b) energy  distribution of neutrinos at
the ISIS beam stop for a  beam current of $I = 200\,\mu$A: \
\numub \ from \mup \ decay (solid), \nueb \ from \mum \ decay
(dashed).\label{fig_nueproduction}}
\end{center}
\end{figure*}
The pulsed spallation neutron source ISIS of the Rutherford
Appleton Laboratory uses  a rapid cycle synchrotron to accelerate
protons up to 800 MeV  with a design beam current of $I = 200 \
\mu$A. The protons are extracted from the synchrotron with a
repetition frequency of 50 Hz as a double pulse, consisting of two
parabolic pulses, with a width of 100 ns and being separated by
325 ns in time. When the 800 MeV protons hit the water cooled
Ta-D$_2$O target $(0.0448\pm0.0030)$ \pip \ per incident proton
are produced \cite{Bur96}. Production of three distinct neutrino
flavours \numu, \nue \ and \numub \ occurs via the \pip -- \mup \
decay chain in the beam stop: \unitlength1.0cm\begin{center}
\begin{picture}(8,2)
\put(0.5,1.7){\makebox(0,0){$\pi^+$}}\put(0.85,1.7){\vector(1,0){0.5}}
\put(1.7,1.7){\makebox(0,0){$\mu^+$}}\put(2.2,1.7){\makebox(0,0){$+$}}
\put(2.7,1.7){\circle{0.6}}\put(2.7,1.7){\makebox(0,0){$\nu_{\mu}$}}\put(6.5,1.7){\makebox(1.5,0)[l]{$\tau_{\pi}
= 26 \; ns$}}\put(1.7,1.4){\vector(0,-1){0.6}}
\put(1.7,0.5){\makebox(0,0){$\mu^+$}}\put(2.0,0.5){\vector(1,0){0.5}}
\put(3.0,0.5){\makebox(0,0){$e^+ +$}}\put(3.7,0.5){\circle{0.6}}
\put(3.7,0.5){\makebox(0,0){$\nu_e$}}\put(4.2,0.5){\makebox(0,0){$+$}}
\put(4.7,0.5){\circle{0.6}}\put(4.7,0.5){\makebox(0,0){$\bar{\nu}_{\mu}$}}\put(6.5,0.5){\makebox(1.5,0)[l]{$\tau_{\mu}
= 2.2 \; \mu s$}} \end{picture}
\end{center}  The  \pip \ and \mup
\ are stopped within the heavy target and decay at rest.
 The unique time
structure of the ISIS proton pulse allows a clear separation of
\numu \ induced events from \numub \ and \nue \ induced events.
 Due to the short life time of \pip \ ($\tau=26$ ns) the \numu \
production closely follows the ISIS proton beam profile. One
therefore expects two \numu \ bursts within the first 600 ns after
the extraction of the proton beam. The 2-body decay at rest of
\pip\ leads to monoenergetic \numu \ with an energy of
$E_{\numu}=29.8$\,MeV. Studies of these \numu \ are published in
\cite{Kncnumu}. On the other hand the \numub \ and \nue \ from
\mup \ decay are expected to emerge on a time scale of a few \us \
due to the \mup \ life time of $\tau=2.2$\,\mus. The time spectrum
of \numub \ and \nue \ induced events [see Fig.
\ref{fig_nueproduction}~(a)] reflects the life time of \mup \  and
thus contains additional information to discriminate in the data
analysis versus background reactions. The \numub \ and \nue \ from
muon decay have continuous energy spectra (see
Fig.\ref{fig_nueproduction}). The energy spectra are well defined
and can be calculated precisely because of the decay at rest
kinematics and the simple V--A structure of the \mup\ decay. From
the three neutrino flavours, which are produced with equal
intensity and emitted isotropically, the highest mean energy is
obtained by the \numub, which have the maximum intensity at the endpoint energy of 52.8 MeV.\\

The intrinsic contamination of the ISIS $\nu$--beam with \nueb \
is very small. The suppression of \nueb --production follows from
the following factors: The stopping of 800 MeV protons in the
Ta-D$_2$O target produces less \pim \  than \pip \
(\pim/\pip=0.56). While \pim , which are stopped quickly
($<1$\,ns), mainly undergo nuclear capture, it is only a fraction
of 1.2\% which decay in flight and therefore become of relevance
for the \nueb \ contamination. The following \mum \ decay at rest
in the target station again is suppressed by the efficient muon
capture ($93\%$ of \mum \ produced) on the high Z material of the
spallation target. This \pim--\mum \ decay chain leads to a very
small contamination of $\nueb / \numub = 6.4\cdot 10^{-4}$
\cite{Bur96} with  the distributions for \nueb \ in energy and
time  shown as dashed lines in Fig. \ref{fig_nueproduction}. The
intrinsic \nueb \ contamination is discussed in more  detail in
section \ref{subsec_konta}. The small \nueb \ component in the
ISIS $\nu$--beam together with the unique time structure of the
proton beam allows a high sensitivity search for \numubnueb \
oscillations.

\subsection{The KARMEN detector}
\begin{figure*}
\begin{center}
\includegraphics{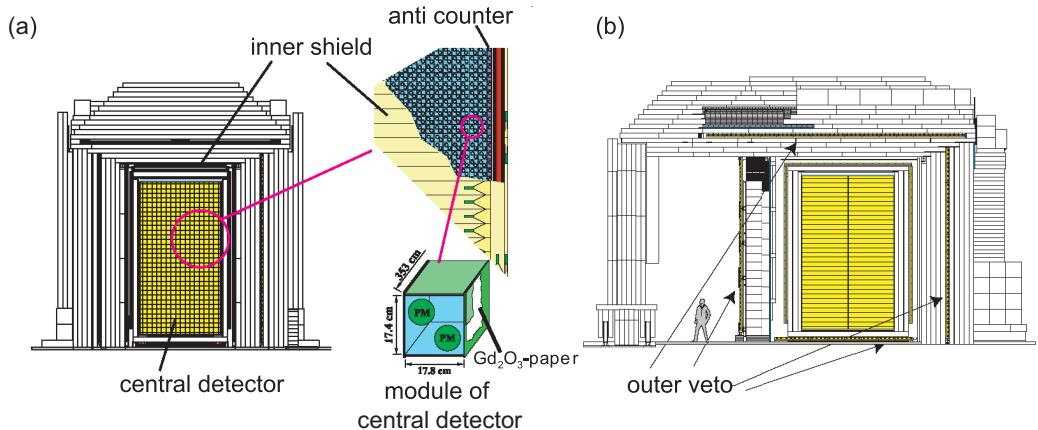}
\caption{ (a) Front view of the KARMEN detector with details of
the central detector region and a single module. (b) Side view,
the ISIS target is located to the right.} \label{fig_det}
\end{center}
\end{figure*}
The KARMEN detector \cite{Dre90} is a segmented high resolution
liquid scintillation calorimeter, located at a mean distance of
17.7 m from the ISIS target at an angle 100 degrees relative to
the proton beam. The liquid scintillator is enclosed by a
multilayer active veto system and a 7000 t steel shielding (see
Fig. \ref{fig_det}). The hydrocarbon acts as active target for
neutrino-nucleus reactions (\C,\Cd,\Hn). The 65 $\mbox{m}^3$ of
liquid scintillator consisted of a mixture of paraffin oil
($75\%$vol.), pseudocumene ($25\%$vol.) and 2 g/l of the
scintillating additive  1-phenyl-3-mesityl-2-pyrazoline (PMP).\\
The liquid scintillator volume is optically separated into
independent modules by an optical segmentation of double lucite
sheets. A small air gap between the double lucite sheets of the
segmentation causes optical total reflection and thus a very
efficient transport of scintillation light to the ends of the
modules, where the scintillation light is read out by a pair of
(3" VALVO XP 3462) photomultiplier tubes (PMT). Furthermore,
gadolinium coated paper has been put between the acrylic walls for
an efficient detection of
thermal neutrons.\\
The segmentation consisted of   608 modules in total, which are
placed inside a rectangular tank with the dimensions of $3.53 \
\mbox{m} \times  3.20  \ \mbox{m} \times 5.96 \ \mbox{m}$ in
length, width and height. The central detector consists of the
inner 512 modules (each with the dimensions of $353 \ \mbox{cm}
\times 17.7 \ \mbox{cm} \times 18.1  \ \mbox{cm}$ in length, width
and height), arranged in 32 rows and 16 columns. A surrounding
layer of modules with half the cross section of a central detector
module defines the inner anti counter. An inner passive shielding
of 18 cm thick steel slabs surrounds the scintillator tank
providing passive shielding and mechanical stability.
 The second layer of
active shielding (inner veto) consists of 136 plastic scintillator
bars (NE110) with thicknes of 3 cm and lengths  ranging from 2.4 m
to 3.1 m, which are mounted onto the passive shielding
on all sides but the bottom side. \\
The surrounding steel shielding is built in a modular way out of
layers of steel slabs. This structure of layers allowed the
integration of an outer veto system inside the steel shielding. In
total, 136 bars of plastic scintillator (Bicron BC412) have been
used for the outer veto system, which
provided also active shielding under the detector. \\
This additional outer veto system was  installed in 1996, marking
the beginning of the KARMEN 2 experiment. The upgrade of the
experimental configuration reduced considerably the background
level for the \numubnueb \ search, as it will be outlined in
section \ref{subsec_cosmic}.

The KARMEN detector is a liquid scintillator calorimeter,
optimized for high energy resolution of $\sigma_E = 11.5\% /
\sqrt{E(\mbox{MeV})}$. An event information comprises  the energy,
time and position information, as well as the number of addressed
modules and their relative time differences.  A scintillator
module hit is accepted if there is a coincidence of signals of the
photomultipliers at both ends within a coincidence time of $\Delta
T_{C1}=190$ ns (first level trigger). The position of the event
along the module axis (x-direction)  is obtained by the time
difference between the signals, whereas the energy information is
derived from the integrated PMT pulses.  The absolute energy
calibration of the detector is fixed by the analysis of the Michel
energy spectrum of electrons from the decay of stopped cosmic ray
muons. The energy calibration is performed for each single module
and takes into account the individual light output curves of the
modules. Module hits within a coincidence time $\Delta T_C<90$ ns
are combined to one event. Analysis of throughgoing muons allow to
calibrate the relative times
 of module hits $t_{rel.}$ with an accuracy of $\delta
 t_{rel.}=0.8 \
\mbox{ns}\ (\mbox{FWHM})$. In the case of events with more then
one module hit, the 3-dimensional position information (x,y,z)
corresponding to module axis, row and column is constructed by the
energy weighted average of the single module information. Finally,
the event time $t$ relative to the ISIS proton beam is recorded.
Individual KARMEN modules are synchronized to the ISIS beam with
an accuracy of $\delta t<2 \ \mbox{ns}$, allowing to exploit the
ISIS time structure in detail. A beam reference time of t=0 is
attributed to the time, when the first neutrino enters the KARMEN
detector. A full description of the detector energy and timing
calibration is given in \cite{Kcalib}.

\section{Oscillation signature \label{sec_signal}}
Neutrino flavor oscillations occur, if the weak interaction
eigenstates \nue , \numu\ and \nutau\ are a superposition of the
non-degenerate mass eigenstates $\nu_1$, $\nu_2$ and $\nu_3$. As
the mass eigenstates propagate differently, there is a non-zero
probability that a neutrino flavor produced via the weak
interaction (e.g. \numub) is detected as another neutrino flavor
(e.g. \nueb) after a traveling distance $L$. In general, the
formalism of the mixing of three flavor and mass eigenstates
requires a unitary $3\times3$ mixing matrix $U$, often referred to
as the Maki-Nakagawa-Sakata \cite{MNS} matrix $U_{\rm MNS}$.
However, the current results in the field of neutrino oscillations
suggest a ´one-mass-scale´ dominance $\delta m^2\equiv \Dm_{12}\ll
\Dm_{13}$ and $\Dm_{13}\approx \Dm_{23}\equiv \Dm$ with $\Dm_{ij}
= |m^2_i - m^2_j|; i,j=1,\dots ,3$
\cite{fogli,bilenky,babu,torrente,minakata,barshay,cardall}.
Possible mixing to sterile neutrinos as suggested by
\cite{suematsu,bereshiani,foot} is ignored whereas CP conservation
is assumed, as we shall do in the following. In this case, and
since the KARMEN experiment with its distance between neutrino
source and detection point of $L\approx 17$\,m is a typical ´short
baseline´ oscillation experiment, it is sufficient to simplify the
mixing scheme to a $2\times 2$ mixing. In such a two flavor mixing
scheme, the probability $P$ to detect a \nueb\ in an initially
pure \numub\ beam with energy $E$ (in MeV) after a path length $L$
(in meters) can be described as:
\begin{center}
\begin{eqnarray}
P(\numubnueb) =  A \cdot \sin^2 \left( \frac{1.27\cdot \Dm \cdot
L}{E}\right) \label{osciprob}
\end{eqnarray}
\end{center}
In a ´short baseline´ regime ($1/\Dm \approx L/E \ll 1/\delta
m^2$), contributions to the oscillation probability $P$ due to the
smaller difference of the squared $\nu$-masses, $\delta m$, can be
neglected. The oscillation amplitude $A$ in (\ref{osciprob}) is a
function of the elements of the mixing matrix $U_{\rm MNS}$. For
simplicity, we define
\begin{center}
\begin{eqnarray}
 A = \sit \label{osciamp}
\end{eqnarray}
\end{center}
keeping in mind, that for a comparison of oscillation searches in
a different mode than \numubnueb\ appearance, one has to calculate
$A$ as the complete function of the $3\times 3$ mixing matrix
elements. For a review on neutrino masses and mixing and a
complete formalism of neutrino oscillations see \cite{Kayser}.

\subsection{\nueb \ absorption on protons}
Appearance of \nueb\ from \numubnueb \ flavor oscillations is
detected by the classical inverse beta-decay on the  free protons
of the scintillator:
\begin{center}
\unitlength0.66mm
\begin{picture}(120,35)
\put(0,26.0){\makebox(10,10){\nueb}}
\put(10,26.5){\makebox(5,10){+}} \put(15,25.7){\makebox(10,10){p}}
\thicklines \put(27,31.5){\vector(1,0){11}}
\put(40,26.0){\makebox(10,10){ n }}
\put(50,26.5){\makebox(5,10){+}} \put(55,26.6){\makebox(10,10){
\pos }} \put(100,26.5){\makebox(10,10){ Q=$-$\,1.804\,MeV }}
\thinlines \put(60,32.5){\circle{8}} \thicklines
\put(45,28){\vector(0,-1){4}} \put(44.5,16.5){\makebox(10,10){
n$_{therm}$ }} \thicklines \put(45,15){\line(0,1){4}} \thicklines
\put(45,15){\line(0,-1){9.5}} \thicklines
\put(45,15.5){\vector(1,0){10}} \thicklines
\put(45,5.5){\vector(1,0){10}} \put(55,10.5){\makebox(5,10){+}}
\put(60,11){\makebox(10,10){ $^{1}$H }} \thicklines
\put(69,15.5){\vector(1,0){5}} \put(76,11){\makebox(10,10){
$^{2}$H }} \put(84,10.5){\makebox(5,10){+}}
\put(89,9.5){\makebox(10,11){$\gamma$}} \thinlines
\put(94,15.5){\circle{8}} \put(55,0.5){\makebox(5,10){+}}
\put(60,1){\makebox(10,10){Gd }} \thicklines
\put(69,5.5){\vector(1,0){5}} \put(76,1){\makebox(10,10){Gd }}
\put(84,0.5){\makebox(5,10){+}} \put(89,0.5){\makebox(5,10){$n$}}
\put(94,0.5){\makebox(10,9){$\gamma$}} \thinlines
\put(99,5.5){\circle{8}} \put(112,0.5){\makebox(10,10){$\langle n
\rangle =3$}}
\end{picture}
\end{center}
%
The \nueb \ signature  is therefore a spatially correlated delayed
coincidence between a prompt positron and a delayed $\gamma$ event
from a $(n,\gamma)$ neutron capture reaction.

\subsubsection{Positron signal \label{sec_posi}}
For different sets of parameters \sit \ and \Dm \ the oscillation
probability $P(\numubnueb)$ is calculated varying \numub \
energies and  flightpaths. These \nueb \ energy spectra are then
transformed into positron spectra by means of the calculated
energy dependence of the \nuebanp \ cross section. The
calculation used \cite{Bea00}  takes into account weak magnetism
and recoil effects, yielding a flux averaged cross section of
$\sigma_{tot} =93.5 \times 10^{-42}$\,cm$^2$ for the \numub \
spectrum from \mup --decay at rest.
\begin{figure*}
\begin{center}
\includegraphics{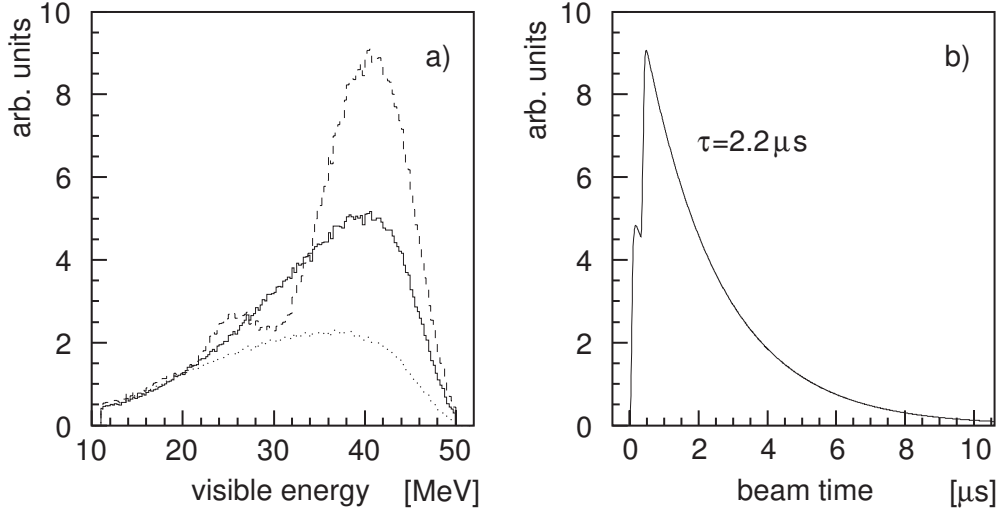}
\caption{Expected \pos \ signal from  \nuebanp . (a) Visible
energy assuming $\Dm=1 \ \eV$  (dotted), $10 \ \eV$  (dashed),
$100 \ \eV$  (solid) and (b) detection time.} \label{fig_nuebsig}
\end{center}
\end{figure*}
Due to the short baseline of $\langle L\rangle$=17.7 m,  the
strongest \numubnueb\ signal is expected at $\Dm =2.8$\,\eV .
Figure~\ref{fig_nuebsig}(a) shows the dependence of expected \pos\
energy spectra for three mass difference values ($\Dm =
1,10,100$\,\eV ), illustrating the modification of the energy
spectrum due to oscillation effects. The  spectra include
experimental  response functions such as energy and spatial
resolutions, threshold efficiencies as well the integration of the
oscillation probability over the detector volume. The visible
energies of positrons extend up to 50\,MeV with the oscillation
signal mostly above 20\,MeV. Figure~\ref{fig_nuebsig}(a) also
demonstrates the power of the detector to discriminate between
different values of \Dm\ in case of a positive oscillation signal.

Apart from the well defined energy spectrum, the  time spectrum of
\pos\ [see Fig.~\ref{fig_nuebsig}(b)], resulting from the unique
ISIS time structure,  discriminates against beam uncorrelated
background. The time distribution of the positrons follows the
2.2\,\us\ exponential decrease of the \mup \ decay at rest. The
positrons are therefore expected in a narrow time window of
several \us\ after beam-on-target .

\subsubsection{Neutron capture signal \label{sec_ngamma}}
\begin{figure*}
\begin{center}
\includegraphics{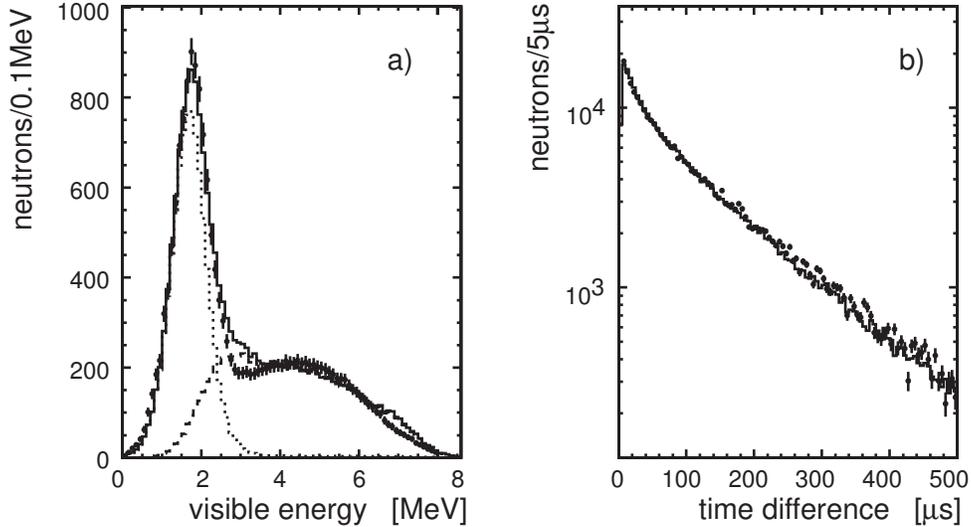}
\caption{(a) Energy and (b) time distribution of neutron capture
  events. The energy signal (experimental data points) is the sum of
  \pnd\ (MC dotted line) and \Gdng\ (MC dashed line) capture. The time
  between neutron production and capture is quasi-exponential with
  a time constant of $\tau \approx 120$\,\us\ well reproduced by MC.} \label{ngsig}
\end{center}
\end{figure*}

The delayed event of the \nueb \ induced delayed coincidence
arises from one of two different neutron capture reactions.
Neutrons from \nuebanp \ reactions have kinematic energies up to 5
MeV and are quickly thermalized. After thermalization, neutrons
are captured either on   protons of the scintillator \pnd\ or on
gadolinium \Gdng\ , which is contained inside the walls of the
segmentation. In the first case, a single mono-energetic 2.2\,MeV
gamma is produced, in the latter case, a complex gamma cascade is
initiated with a sum energy of $\sum E_{\gamma} = 7.9~\mbox{MeV}$
\footnote{The n-capture on $^{157}$Gd with a cross section of
$\sigma=254000\cdot 10^{-24}\mbox{cm}^2$  and an endpoint energy
of $E_0=7937.4$~keV \cite{Whi} dominates over the n-capture on
other Gd-isotopes.
} [see Fig.~\ref{ngsig}(a)].\\
Neutron capture reactions are monitored {\it in situ} during the
measurements by  investigating the capture reaction
\begin{equation}\label{eq_mucap}
 \mucapxn
\end{equation}
of stopped cosmic ray muons. This reaction produces neutrons with
kinetic energies in the few MeV range \cite{cosNE}, comparable to
the energy of neutrons from the \CCprot\ process.
Figure~\ref{ngsig}(a) shows the measured spectrum of visible
energies following a stopped muon in a coincidence volume of
$V_c\approx1 \ \mbox{m}^3$ ($|\Delta x|<60$ cm, $|\Delta{\rm
row}|,|\Delta{\rm col}|\le 2.5$) around the endpoint of the muon
track. The \pnd\ peak can be clearly separated from the broad
distribution of \Gdng\ signals. The \Gdng \ signal does not peak
at $E_0=7.9 \ \mbox{MeV}$ due to the calorimetric properties of
the single modules. If the $\gamma$'s from the cascade are spread
over different modules, missing visible energy can occur due to
the thresholds of
individual modules. \\
The neutron thermalization and capture followed by $\gamma$
emission is simulated using the {\small GEANT/GCALOR} program
\cite{geant,czeit}. The simulated spectra shown in
Fig.~\ref{ngsig}(a) include detector response functions and have
been adjusted separately to the measured distribution. For visible
energies below 3--4\,MeV the energy resolution,  as well as
hardware thresholds together with the complex topology of a
multi-$\gamma$ event lead to difficulties in describing the
spectral shape by Monte Carlo simulations. However, since \mum \
capture reactions (Eq. \ref{eq_mucap}) are measured, the spectral
shape of neutron capture events  and the total neutron detection
efficiency can be reliably
measured, in order to be used for the \numubnueb \ search. \\
The experimental as well as the MC generated  time difference
between the prompt cosmic muon and the $\gamma$'s from the neutron
capture is shown in Fig.~\ref{ngsig}(b). The distribution can be
approximated by a single time constant of $\tau \approx 120$\,\us
, reflecting the thermalization and diffusion processes of the
neutron and the subsequent two competing capture processes. There
is a slightly enhanced occurence of $\gamma$'s within the first
\us \ is due to a higher rate of \Gdng \ capture. This is
explained by the almost immediate capture of neutrons being
produced near the walls  containing Gd.

\subsubsection{Neutron detection efficiency \label{sec_neff}}
The neutron detection efficiency  $\varepsilon_{N}$ has to be
determined accurately in order to calculate the expected number of
(\pos ,n) sequences from \numubnueb\ oscillations. The efficiency
$\varepsilon_{N}$ is determined by monitoring the nuclear capture
reactions of stopped muons (Eq. \ref{eq_mucap}). It is given by
the ratio of detected neutrons $N_n$ to the
total number of produced neutrons $M_n$. \\
 The number of
detected neutrons $N_n$ is given by the number of delayed
coincidences occuring after a stopped muon. According to the
expected neutron capture signal, we require the delayed event to
occur within a coincidence time  $5\le\Delta t\le 300\,\us$ with
energies $E_{del.}\le 8$\,MeV and within a coincidence volume of
$V_c= 1.3\  \mbox{m}^3$.\\
In order to derive the total number of produced neutrons $M_n$,
the number $N_{\mu^-}$ of stopped \mum, the \mum\,capture rate
$\Lambda_c$, and the neutron emission multiplicity $\langle x
\rangle$  must be known.
 As the charge of stopped cosmic muons
 cannot be determined for individual tracks, the decay time
spectrum has been analyzed to derive the charge ratio
$\mup/\mum=R_\mu=1.28\pm0.03$ and thus the number $N_{\mu^-}$ of
stopped \mum \ is known from the measured number of stopped muons
$N_\mu$. With a total \mum\ capture rate of
$\Lambda_c^{tot}=(38.4\pm0.4)\cdot 10^{-3}$\,s$^{-1}$ on \C\
\cite{suzu} corrected for the abundance of \Cd\ and \Og\ in the
scintillator, an average probability per stopped \mum\ of
 $\alpha_c^n=(64.1\pm1.3)\cdot 10^{-3}$ is derived for processes with neutron
production.

The derived neutron detection efficiency $\widetilde{\varepsilon}$
from these values
\begin{equation}\label{eq_neffcalc}
\widetilde{\varepsilon}= \frac{N_n\cdot (1+R_\mu)}{N_\mu \cdot
\alpha_c^n}
\end{equation}
must then be modified in two aspects: \\
(1) Due to  multiple neutron emission $\langle x \rangle=1.07$
(see Eq. \ref{eq_mucap}), the derived efficiency
$\widetilde{\varepsilon}$ must be corrected to the single neutron
expectation from the \nuebanp \ reaction.\\
(2) As the identification of the muon stop point can lead to
ambiguities for tracks, which stop close to the borders of the
detector, a restricted fiducial volume of the detector to the stop
points of muons ($|x_{stop}|<150$\,cm, the outermost module layer
removed) is applied. The detection efficiency
$\widetilde{\varepsilon} $ is then extrapolated to the entire
detector volume using  {\small GEANT/GCALOR} simulations.

\begin{figure*}
\begin{center}
\includegraphics{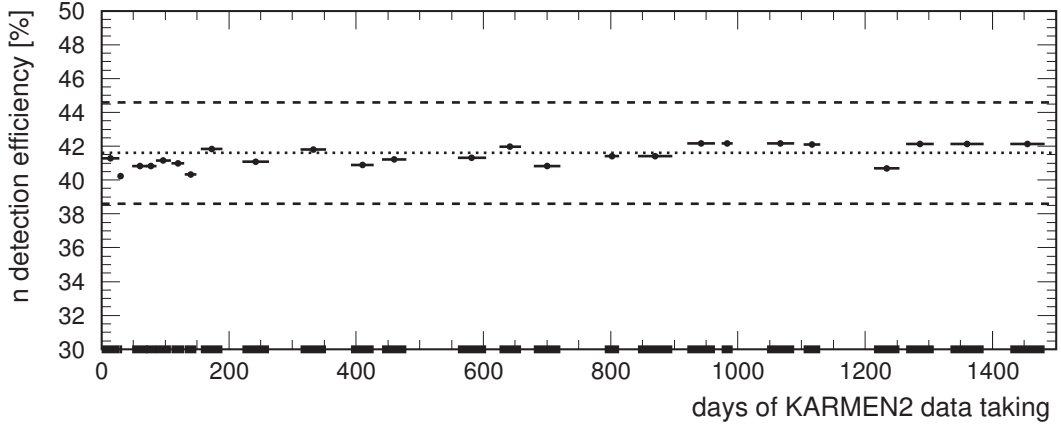}
\caption{Measured single neutron detection efficiency as a
function of time
  during data taking. The horizontal bars indicate ISIS beam-on intervals,
  the dotted line shows the neutrino-flux weighted average of the neutron
  detection efficiency, the dashed lines the total systematic error band.} \label{n-eff}
\end{center}
\end{figure*}

A complete description of the analysis of muon capture reactions
with the KARMEN detector and the derivation of the neutron
detection efficiency
is given in \cite{janna}. \\
 Taking all
effects into account, the neutron detection efficiency
$\varepsilon_{N}$ amounts to:
\begin{equation}\label{eq_neff}
\varepsilon_{N}=0.42\pm 0.03
\end{equation}
This value is the neutrino flux weighted average of the entire
KARMEN\,2 measuring period as shown in Fig.~\ref{n-eff}.
\subsection{\nueb\ absorption on carbon} 
A second   \nueb\, detection reaction is the inverse
beta decay of carbon \CBn\ 
with a Q-value of 16.7 MeV. This \nueb\ detection reaction has a
smaller flux-averaged cross section \cite{Kol00}
 than \CCprot . In
addition, the  number of target atoms $N_{T}$ in the scintillator
is  smaller than the number of free protons (see table
\ref{tab_nuebanc}). It thus contributes  about 5\% to the
detection of \nueb. The {\small GEANT 3.21} Monte Carlo simulation
of \nuebanc \ is included in the total number and spectral shape
of expected (\pos ,n) sequences from \numubnueb\ oscillations
[Fig. 3(a)].
\begin{table} [h]
  \centering
  \begin{ruledtabular}
  \begin{tabular}{ l c c}
             &         $\nuebanp$           & $\nuebanc$ \\
\hline
  $N_{T}$ & $ 4.5 \cdot 10^{30}$  & $2.5\cdot 10^{30}$ \\
  $ \sigma(\numubnueb)$ & $ 93.5\cdot 10^{-42} \ \mbox{cm}^2$  &
  $8.5\cdot10^{-42} \ \mbox{cm}^2$ \\
    $ \sigma(\nueb \ \mbox{contamination})$ & $ 72.0\cdot 10^{-42}\  \mbox{cm}^2$  &
  $7.4 \cdot10^{-42} \ \mbox{cm}^2$
  \end{tabular}
  \end{ruledtabular}
  \caption{Comparison of flux averaged cross sections $\sigma$ and target nuclei $N_{T}$ for  detection of \nueb \ from different sources.}\label{tab_nuebanc}
\end{table}

\section{General event requirements \label{subsec_req}}
\begin{figure*}
\begin{center}
\includegraphics{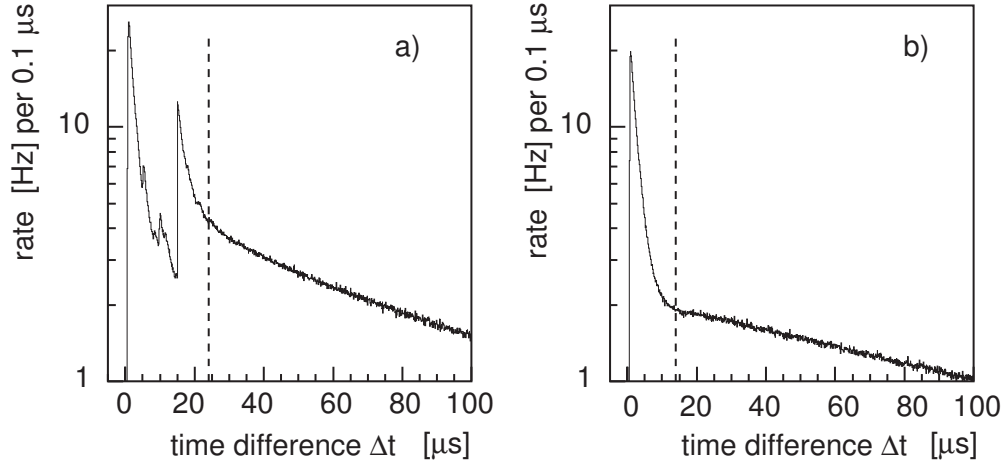}
\caption{Rate of events following in a time difference $\Delta t$
to the last preceding event  (a) in main detector, inner veto or
anti counter (b) in outer veto system. The count rate suppression
for time differences $\Delta t<15 \ \us$  in Fig. (a) is caused by
hardware and software deadtimes as well as read-out dead times.}
\label{fig_stack}
\end{center}
\end{figure*}

The special feature of the \numubnueb \ signature is its delayed
coincidence nature of a prompt high energetic positron, followed
by a low energetic signal from neutron capture. Before enforcing
stringent cuts, which correspond to the delayed coincidence nature
of the \nueb\ detection reaction, we apply loose cuts to the data
set, which do not cut into the signal region but which strongly
suppress background. \\
(1) Only sequences of two events are accepted.\\
(2) A sequence accepted for further evaluation in the software
analysis consists  of a prompt event and a delayed event which
shows the typical characteristics of neutron capture events. In
particular, this means that the delayed signal occurs within
$\Delta t < 500 \ \mus$ \ after the prompt event and has energies
less
than $E_{del.} < 8 \ \mbox{MeV}$. A coincidence volume of $V_c=1.3 \ \mbox{m}^3$ is required. \\
(3) Neither the prompt event nor the delayed event must have
any hits in the multilayer veto system. \\
(4) The prompt event must have energy $E_{pr.}>11$~MeV.\\
(5) There must be no activity in the detector system preceding a
prompt event. The history of all activities in the detector system
(total trigger rate $\Gamma_{tot}\sim13$\,kHz) are stored by a
time stamp and a bit pattern word, which allows the decryption of
addressed detector parts. Requesting no activities preceding an
event in the main detector, inner veto or inner anti counter in
the previous 24 \us \ (14 \us \ for the outer veto system)
eliminates most of the cosmic induced background with
short time correlations, as shown in figure \ref{fig_stack}.\\
(6)  There must be no stopped cosmic ray muons in the central
detector preceding a prompt event. With a rate of
$\Gamma_{\mu}\sim 160 $ Hz the hardware trigger identifies stopped
muon in the central detector. A 10 \us \ hardware dead time is
then applied and
 the event time and stopping position of the muon are
stored, thus providing information for the offline analysis to
detect spatial correlations between an event and preceding stopped
muons. Prompt events of a potential \nueb \ coincidence are
rejected, if they occur within $\Delta t< 40 \ \us$ \ after
stopped muons anywhere in the central detector, after up to
$\Delta t< 500 \ \us$ \ within a coincidence volume of $V_C=1.3 \
\mbox{m}^3$ (\mum\ capture with n-emission),   or if they occur in
a coincidence volume of $V_C=0.5 \
\mbox{m}^3$ for time differences $\Delta t < 100 \ \mbox{ms}$ (\mum\ capture with subsequent \B\, $\beta$~decay).\\
(7) In the case of events with more than one addressed module in
the central detector,  the maximum time difference between the
module
hits must not exceed $\Delta T_{cmod}=50$\,ns, ensuring that the module hits belong to the same physical event.\\
(8) Not more than 10 modules of the central detector must be addressed.\\
\section{Background reactions \label{sec_bg}}
Evidence for flavor oscillations \numubnueb\ in the appearance
mode requires statistically significant detection of \nueb\ in the
time window of \numub\ in excess of any inherent background. While
for maximal mixing one expects several thousand oscillation
events, a mixing amplitude $10^{-3}<A<10^{-1}$ (as suggested by
LSND) could reduce this number to about 10 events. Despite the
clear oscillation signal and the small ISIS duty cycle, the clear
and unambiguous detection of such rare \nueb\ events requires a
very efficient detection and suppression of the large amount of
cosmic induced reactions. Benefiting from the three-fold active
veto system the cosmic background can be suppressed to a level
well below the expected oscillation events.

However, neutrino induced reactions can also induce a background
rate. In particular,  \nue\ induced  charged and neutral current
reactions constitute the largest background reactions in the
search for \numubnueb\ oscillations. This section discusses both
background reactions in the \nueb \ search, induced by cosmic rays
as well as by neutrinos.

\subsection{Cosmic induced background \label{subsec_cosmic}}
The  cosmic ray induced background reactions are measured  in the
long beam--off time window between the $\nu$--pulses. Taking into
account the trigger structure of the experiment, which also allows
for calibration measurements, the effective statistics for cosmic
induced reactions in the beam-off time interval is 140 times
larger than the narrow time interval for the $\nu$--pulse. This
factor allows to extrapolate the determined cosmic induced
background rate with a statistical  accuracy of 5\% of the
neutrino analysis.

The 7000 t steel shielding of the detector absorbs both the
hadronic and electromagnetic component of cosmic rays. It is
therefore only the muonic component, which can induce \nueb--like
background processes.

\subsubsection{Throughgoing muons}
The KARMEN central detector was exposed  to a rate of $1.1$~kHz of
throughgoing muons. These muons were detected  in the central
detector modules, as well as  by the active veto system. The veto
system inefficiency is estimated to be less than $2.2 \times
10^{-5}$. Delayed activities following cosmic ray muons by
spallation processes of high energetic muons on \C, are highly
suppressed due to the general event requirement~5 (see
section~\ref{subsec_req}) and can be neglected in the \nueb\
search.

\subsubsection{Stopped muons}
Stopped muons in the central detector can cause  spatially
correlated events on the time scale of a few microseconds up to
several milliseconds. Whereas all \mup\ stopping in the detector
will decay, a fraction of $\alpha_c=7.8\%$ of the stopped \mum\
undergo nuclear capture reactions in the scintillator. The muon
decay produces a spatially correlated electron or positron with an
energy up to $E_0=52.8\ \mbox{MeV}$. The time correlation is
defined by the lifetime of \mup \ ($\tau= 2.197 \ \us$) and \mum\
($\tau= 2.026\ \us$). With a branching ratio of
$\Gamma_{\mum}=0.82$, the nuclear capture reactions involve
neutron production:
\begin{equation}\label{eq_mucap2}
  \mucapxn
\end{equation}
The neutrons are detected by the typical neutron capture events of
p(n,$\gamma$) or Gd(n,$\gamma$) with $E_0=8$~MeV and
$\tau_{capture}\approx 120 \ \us$. This process leads to a
contribution to the cosmic induced background in the \nueb \
search, which arises from unvetoed muons with short track lengths,
stopping in the central detector and depositing less than 51~MeV.

Long lived background arises from muon capture reactions of \mum\
\begin{equation}\label{eq_mucapBgs}
  \mucap
\end{equation}
 to the \B \ ground state  or $\gamma$--unstable levels, through the subsequent
$\beta$--decay:
\begin{equation}\label{eq_Bdecay}
  \Bdecay
\end{equation}
with $\tau=29.1$~ms and an endpoint energy of $E_0=13.3$\ MeV for
the beta--electron. Hence, this reaction has only a small overlap
in its signature to  \nueb\ induced coincidences. Nevertheless,
each event arising in the main detector is checked for preceding
stopped muons for time differences up to $\Delta t< 100$~ms
(general event requirement no.6) to suppress the beta decay, whose
electrons otherwise give rise to random coincidences.

\subsubsection{Muons near the central detector}
The dominant cosmic ray induced background is due to muon
interactions in the 7000~t steel shielding blockhouse, which
generate highly energetic neutrons. Two different reaction
mechanisms can be distinguished:
\begin{itemize}
\item \mum\ capture on \Fe :
\begin{equation}\label{eq_mucapfe}
\mum \ + ^{56}\mbox{Fe} \ \longrightarrow \ ^{56-x}\mbox{Mn} \ + \
x\cdot n \ +\
 \numu
\end{equation}
 Negative charged muons
 stopped in iron are predominantly captured with a capture rate of
 $\lambda_c=(4.411\pm0.026)\times10^6/s$ \cite{Suz87}. The energy
 transferred to the nucleus in the process is between 15 and 20~MeV
 and therefore above the neutron emission threshold.
\item Deep inelastic scattering (DIS) of muons on \Fe :
\begin{equation}\label{eq_spall}
\mupm \ + ^{56}\mbox{Fe} \ \longrightarrow \ \mbox{X} + y\cdot n +
\mupm
\end{equation}
Virtual photons  radiated from the cosmic muons interact with the
iron nuclei and can produce spallation neutrons with energies up
to a few GeV. On average, 3-4 secondary particles with energies
above 10~MeV are produced, primarily neutrons and protons.
\end{itemize}
\begin{figure*}
\begin{center}
\includegraphics{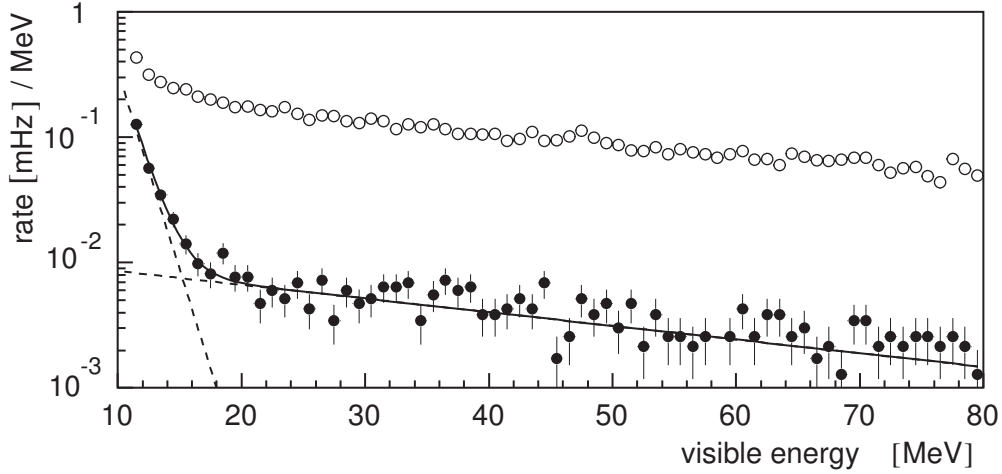}
\caption{Energy distribution of prompt events of cosmic induced
sequences. Measurement ignoring information (open dots) and using
information (full dots) of the outer veto system. See text for
details on the exponential fits.} \label{fig_cbred}
\end{center}
\end{figure*}
Neutrons from deep inelastic scattering can penetrate into the
liquid scintillator, causing signals  with visible energies up to
200~MeV through elastic n--p scattering. After thermalization the
neutrons are captured either on protons or on the gadolinium,
yielding capture $\gamma$ spectra, as shown in Fig.~\ref{ngsig}.
Thus, the highly energetic neutrons cause delayed coincidences,
which are nearly identical to the signature of \nueb, as the
KARMEN detector has no particle identification and cannot
distinguish between cosmic induced n-p recoil events and positrons
from \nuebanp. The crucial identification of the highly energetic
neutrons is achieved by the third veto counter system, which is
placed inside the steel shielding. Figure~\ref{fig_cbred} shows
the spectrum of the visible energies of the prompt events,
covering the entire energy interval of a potential oscillation
signal. The delayed events of these sequences follow the expected
distributions for neutron capture (see fig.~\ref{ngsig}).
\begin{figure*}
\begin{center}
\includegraphics{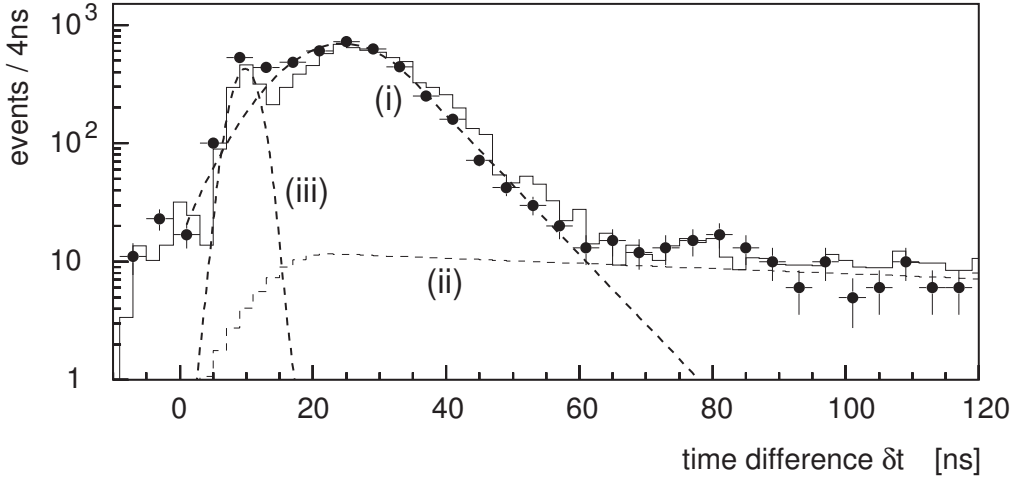}
\caption{Distribution of time difference $\delta t$ between hits
in the outer veto and subsequent hits in the central detector of
cosmic induced background. The Monte Carlo simulation (solid line)
consists of three components: (i) fast neutrons from DIS, (ii)
neutrons from muon capture on iron, (iii) stopped muons.}
\label{fig_trel}
\end{center}
\end{figure*}

Figure \ref{fig_trel} shows the identification of the processes
involved by the time correlation of prompt muons and the proton
recoil event. The time distribution is measured by the time
difference $\delta t$ between the hit in the outer veto system
caused by the muon and the subsequent hit in the central detector
caused by the proton recoil from highly energetic neutron
interaction. The time distribution shows three
components:\\
(i) The dominant gaussian shaped distribution  peaking at a time
difference of $\delta t =25 \ \mbox{ns}$ with an additional
enhanced tail distribution, which  can be attributed to highly
energetic neutrons from deep inelastic muon scattering on iron.
The time difference for these events is equivalent to the time of
flight of the neutrons from their point of production  in the
steel shielding to their n--p interaction in the central
detector.\\
(ii) For time differences $\delta t > 60\ \mbox{ns}$ neutrons from
\mum \ stopping in iron with subsequent nuclear capture \Fen \
dominate. The time correlation of these neutrons largely reflects
the capture rate  of muons in iron $(\tau=206
\ \mbox{ns})$ \cite{Suz87}. \\
(iii) In the time interval $0<\delta t< \ 20 \ \mbox{ns}$ there is
an additional component, caused by muons which hit the outer veto
and stop within the central detector. In this case, the time
distribution corresponds to the muon time of flight from the veto
to the central detector.

The solid histogram in figure~\ref{fig_trel} represents the
expected time distribution from {\small GEANT3.21} simulations,
which are in good agreement with the experimental data and are
described in detail in \cite{Armfzka}.

Having identified events induced by cosmic ray interactions on
iron using the outer veto, this background is strongly suppressed.
The measurement indicated by full circles in
figure~\ref{fig_cbred} shows  the remaining cosmic induced
background, if sequences are rejected where the prompt events have
simultaneously addressed modules in the central detector and in
the outer veto system. These remaining  sequences constitute the
cosmic ray induced background for the \nueb \ analysis. They arise
from the fraction of neutrons, which are produced outside the
outer veto system, and are not absorbed in iron on their path to
the detector (attenuation length of highly energetic neutrons in
iron $\Lambda = 21.6 \ \mbox{cm}$\cite{Bur94}). The remaining
spectrum consists of two components. The soft component is caused
by neutrons from muon capture reactions and can be described as an
exponential distribution $e^{-E/E_0}$ with $E_0\approx 1.4$~MeV.
The much harder component is attributed to neutrons which have
been produced in deep inelastic scattering processes of cosmic ray
muons. This second component with a parameter of $E_0\approx
42$~MeV covers the entire region of interest for the oscillation
search.

Compared to the background rates before the installation of the
outer veto system (corresponding to the energy spectrum with open
circles in fig.~\ref{fig_cbred}), a background suppression by a
factor 35 is achieved, resulting in a total rate of
R$_{CB}=(0.20\pm0.01)$~mHz for the data cuts of the \numubnueb\
analysis in section~\ref{sec_eva}. With this rate the cosmic
induced background is smaller than the neutrino induced
background.

\subsection{Neutrino induced background}

A second source of background reactions arises from the charged
current (CC) and neutral current (NC) interactions of \nue\ and
\numub\ with the carbon nuclei of the liquid scintillator and iron
nuclei of the inner passive shielding. To estimate the background
contributions arising from different CC and NC reaction channels,
the experiment takes advantage of having measured all relevant
cross sections in a series of precision measurements
\cite{K_cc,Kncnue}. Thus, the calculated number of background
events from conventional neutrino interactions does not rely on
theoretical estimates of neutrino induced cross sections. This is
especially important, as the $\nu$-induced background is the
dominant background contribution to the KARMEN neutrino
oscillation search.

In the following we discuss the different $\nu$-induced background
reactions in detail. For each background component we specify the
experimental cross section as well as the detailed spectral
information on energy and time, which have been used to calculate
its contribution to the \numubnueb\ oscillation search.

\subsubsection{The \nue\ induced charged current reaction \label{subsec_excl}}
\begin{figure*}
\begin{center}
\includegraphics{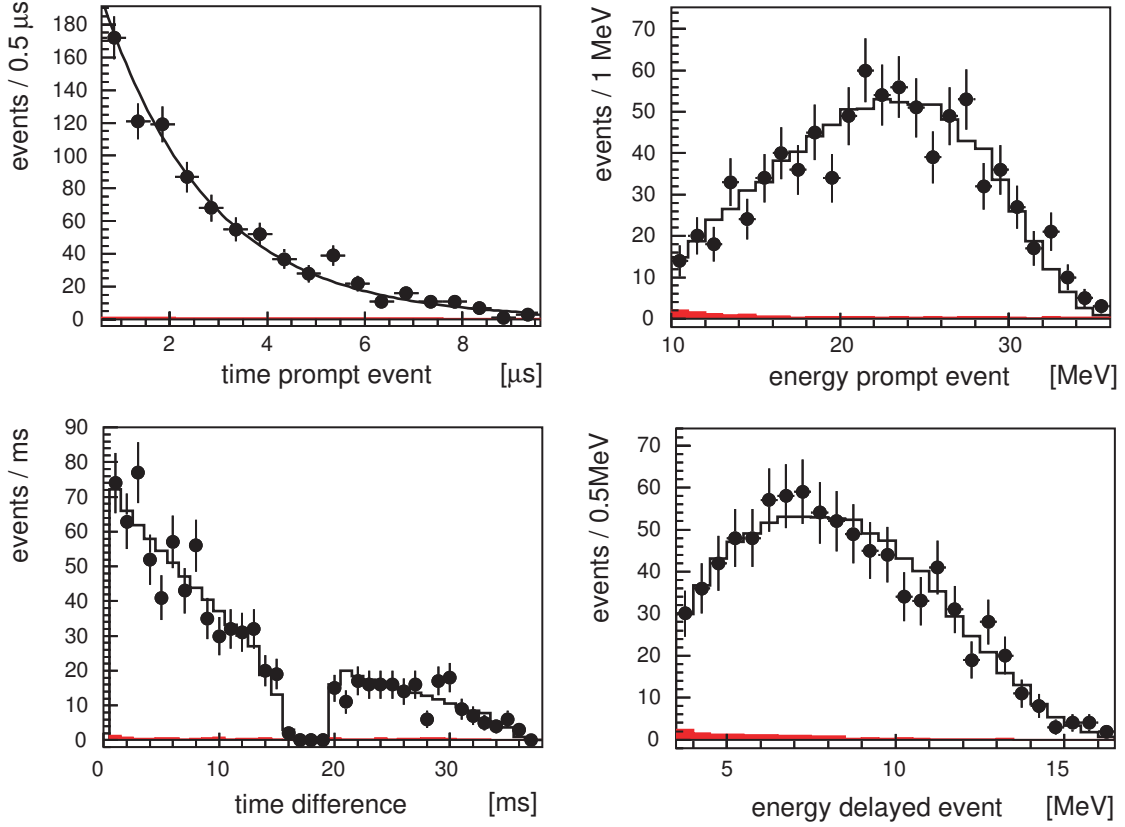}
\caption{Measurement of \excl \ reactions (measuring points),
leading to long lived coincidences between prompt \el \ and
delayed \pos \ from \Ngs \ decay (solid line (MC), shaded area
(background)). (a) event time of \el , (b) visible energy of \el,
(c) time difference between \el \ and \pos , (d) visible energy of
\pos.} \label{fig_cc}
\end{center}
\end{figure*}
Exclusive charged current interactions of \nue \ with \C\ can be
detected by a delayed coincidence consisting of a prompt electron
from  the inverse beta reaction \excl \ and the subsequent
detection of a delayed positron from \Ngs \ decays:
\begin{center}
\unitlength0.8mm
\begin{picture}(120,10)
\put(0,6.0){\makebox(10,10){\nue}} \put(10,6.5){\makebox(5,10){+}}
\put(15,7.0){\makebox(10,10){\C}} \thicklines
\put(27,11.5){\vector(1,0){11}}
\put(40,6.5){\makebox(10,10){\Nzg}}
\put(50,6.5){\makebox(5,10){+}} \put(55,6.5){\makebox(10,10){\el}}
\thinlines \put(60,11.5){\circle{8}} \thicklines
\put(45,7){\line(0,-1){5.5}} \thicklines
\put(45,1.5){\vector(1,0){10}} \put(55,-3.5){\makebox(10,10){\C}}
\put(65,-3.5){\makebox(5,10){+}}
\put(70,-3.5){\makebox(10,10){\pos}} \thinlines
\put(75,1.5){\circle{8}} \put(80,-3.5){\makebox(5,10){+}}
\put(85,-3.5){\makebox(10,10){\nue}}
\end{picture}
\end{center}
The lifetime of \Ngs \ is $\tau=15.9 \ \mbox{ms}$ and the
$\beta$--decay endpoint amounts to $E_0=16.3 \ \mbox{MeV}$. In
total, 860 sequences of this type have been identified with a
signal to background ratio of 61:1. Figure~\ref{fig_cc} shows the
spectral information of the measured sequences, which are both for
the prompt events and the delayed events in good agreement with
the expectation from simulation. This fact underlines the
reliability of the use of these simulated spectra in the
likelihood analysis later applied. The measurements of KARMEN\,1
and KARMEN\,2 show full  compatibility. For definiteness we use in
the following the published CC event sample of KARMEN\,1
\cite{K_omega}, which leads to a cross section of
\begin{equation}\label{eq_cscc}
\sigma=[9.4\pm0.4(\mbox{stat.})\pm0.8(\mbox{sys.})]\times10^{-42}\mbox{cm}^2
\end{equation}
It is the  small fraction of 1.7\% of \Ngs \ decaying within the
first $300 \ \us$ \ and  depositing visible energies of less than
8~MeV which contribute to the expected background in the \nueb \
search. This background is extrapolated  from the measured number
of charged current sequences with time differences of $0.5<\Delta
t<35.5 \ \mbox{ms}$ to the smaller time interval $0.5<\Delta t<300
\ \us$ on basis of the known \Ngs \ lifetime and the \Ngs \ energy
spectrum. The uncertainties in the extrapolation correspond to 5\%
accuracy in the prediction of this background component.

Charged current reactions of \nue \ on iron with subsequent
neutron evaporation from the excited iron nucleus \CCFen \ have
been investigated and simulated as possible background channel.
Despite the rather high cross section calculation of $\sigma =
34.8\times 10^{-42}\mbox{cm}^2$ for this reaction channel
\cite{Kol99} and the significant number of target nuclei of the
inner passive shielding $N_T=2.4\times10^{30}$, \nue \ reactions
on iron do not give rise to background in the \numubnueb \
analysis. The suppression of this channel is caused by the low
efficiency of the electrons, which are produced with energies up
to 35~MeV inside the steel, to be detected in coincidence with the
neutron events inside the central detector.

\subsubsection{Random coincidences}
\begin{figure*}
\begin{center}
\includegraphics{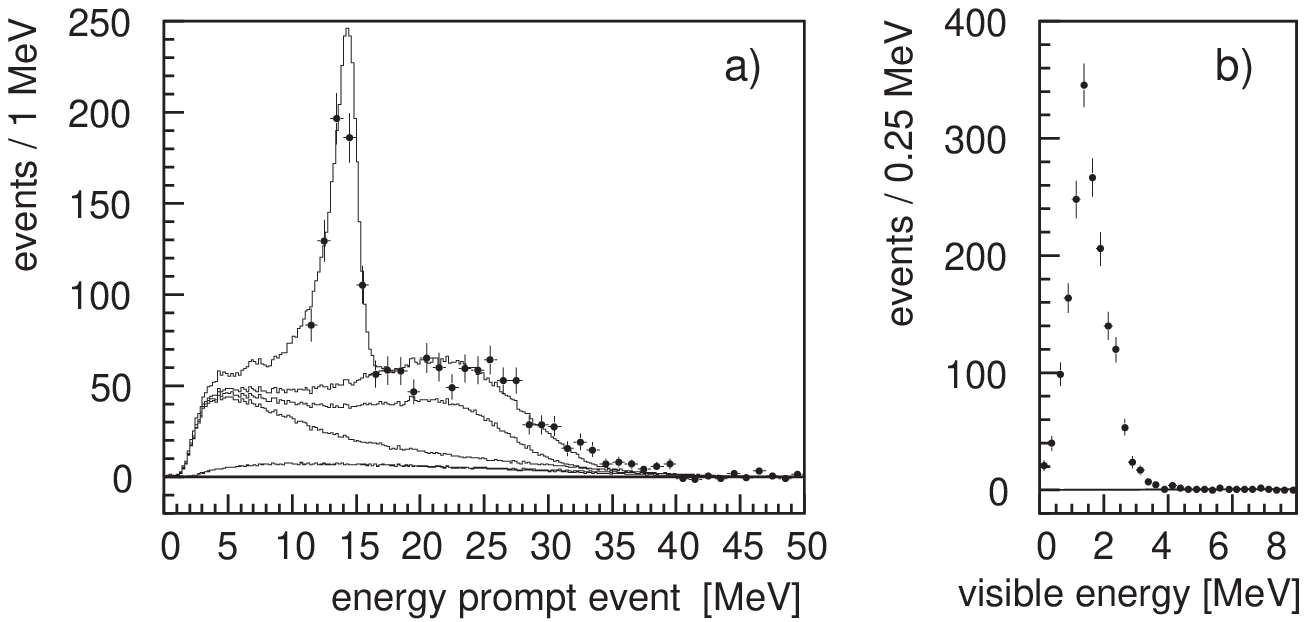}
\caption{(a) Measurement of neutrino induced reactions in the time
window $0.6<t<10.6 \ \us$. The calculated contributions are
(bottom to top): $\nu-\el $ scattering, \CCFe , \CCexc, \excl and
\NC . (b) Energy distribution of  uncorrelated delayed events. }
\label{fig_rnd}
\end{center}
\end{figure*}
Neutrino-nucleus interactions as well as neutrino electron
scattering increases the number of events in the positron time
window in the first few \us \ after beam on target. This implies
an enhanced rate of random coincidences between a neutrino induced
(prompt) event and a low energy event  from environmental
activity. Random coincidences, initiated by cosmic induced events,
are accounted for in the measurement of the cosmic induced
background. The probability $P_{rc}$ of  an uncorrelated event to
follow a neutrino-nucleus interaction as well as its spectral
information are extracted by applying the search criteria to
uncorrelated events, for example to events recorded in preceding
beam periods. This method allows to determine  $P_{rc}$ and the
spectral information of the delayed events with high statistics.

In the energy range from $11<E<50$~MeV and in the time window
$0.6<t<10.6 \ \us$  after beam on target, 1567 neutrino-nucleus
interactions are measured. The  neutrino interactions arise mainly
from two different types of neutrino-nucleus interactions. The
largest contribution arises from the inclusive charged current
reaction \CC , as well as from  neutral current reaction \isov\
with $\nu =(\nue,\numub)$.  The neutral and charged current
contribution are clearly visible in the energy spectrum of the
measured neutrino-nucleus interactions (see Fig.~\ref{fig_rnd}).
The delayed events of random coincidences are uniformly
distributed in time, and their energies are close to the threshold
of a single detector module (mean energy $ \langle E \rangle
=1.1$~MeV) as shown in figure~\ref{fig_rnd}(b).

The probability $P_{rc}$ for an uncorrelated event to occur with a
time difference of up to $5<\Delta t<300$~\us\ and within a
coincidence volume of $V_c=1.3 \mbox{m}^3$ after the prompt event
is determined to be $P_{rc}=(5.5\pm0.4)\times10^{-3}$
\footnote{This number is valid for the final data cuts given in
section~\ref{sec_eva}.}.

The expectation value for the neutrino-induced random background
$N_{rc}$ is obtained by multiplying the number of measured
neutrino-induced reactions $N_\nu$ with the  probability $P_{rc}$.
Using this method, the statistical accuracy of $N_{rc}$ is $7\%$.
The measured spectral information is used for the likelihood
analysis.

\subsubsection{\nueb\  contamination \label{subsec_konta}}

The only background source, which can not be directly extracted
from the data, is the contamination of the neutrino beam with
\nueb \ produced in the \pim -- \mum\ decay chain. Detailed Monte
Carlo simulations, including a three-dimensional model of the ISIS
target,  and its surroundings are used to obtain the fraction of
\pim\ and \mum\ decaying before they undergo  capture on nuclei of
the target materials \cite{Bur95,Bur96}. The overall ratio of
\nueb \ produced  in the ISIS target relative to \numub \ from
\mup--decay amounts to $\varepsilon=6.4\times10^{-4}$. This ratio
is further reduced by taking into account, that  the lifetime of
\mum \ depends on the target material and is in general shorter
than the \mup \ decay time (see figure~\ref{fig_nueproduction}),
leading to a further reduction of \nueb \ by a factor of 0.764 in
the time window of $0.6<t<10.6 \ \us$. Finally, the \nueb \
spectrum from \mum \ decay (Fig.~\ref{fig_nueproduction}(b)) leads
to a lower flux averaged cross section of $\sigma = 72.0\times
10^{-42}\mbox{cm}^2$ for the \nuebanp\ reaction (see
table~\ref{tab_nuebanc}). Taking all effects into account, the
intrinsic \nueb \ contamination leads to the smallest background
contribution in the \nueb \ search.

\subsection{Beam correlated neutron background \label{subsec:br_neutrons}}

Each 800~MeV protons of the ISIS beam produces typically  25
spallation neutrons in the target with energies up to 400~MeV
\cite{spall}. The 7~m steel shielding between ISIS target and
detector reduces  the neutron flux by a factor of more than
10$^{15}$. Despite the flux reduction, punch--through neutrons are
observed in the central detector. However, these high energy
neutrons closely follow the ISIS double proton pulses
\cite{Kncnumu} and are restricted to the time window of $t<500 \
\mbox{ns}$ after beam on target. Setting the lower time cut for
the positron window at $t_{pr.}>600$ ns after beam on target,
completely eliminates reactions from these neutrons.
\section{Data reduction \label{sec_eva}}

\subsection{Raw data}

The results presented here are based on measurements from February
1997 to March 2001. During this time, protons equivalent to an
accumulated total charge of 9425 Coulombs have been stopped in the
ISIS target. This corresponds to a total number of
\begin{equation}
N_{\nu}=2.71\times 10^{21} \end{equation} neutrinos for each of
the flavors \nue , \numub\ and \numu\ produced at the ISIS beam
stop.

In total, the KARMEN data acquisition system recorded
$3.7\times10^9$ events. Out of these single events,
$1.93\times10^7$ have no hits in the veto counter system and
deposit more than 11~MeV and hence can be classified as candidates
for a prompt event of a delayed coincidence. Requiring in addition
the detection of a second event without  veto hits in the
following  500 \us \, results in $3.5\times10^5$ delayed
coincidences. After application of the general event requirements,
defined in section \ref{subsec_req}, the sample size shrinks to
3464 coincidences with  more than 99\% of these coincidences
outside the time window of the \numubnueb \ analysis.

The detector system was 777.4 days online, excluding additional
measurements for specific background studies and calibration
purposes. Taking into account ISIS  beam on times, the duty cycle,
and a 10~\us\ long neutrino time window, the effective neutrino
measuring time amounts to 7.5 hours.

\subsection{ Final selection criteria \label{subsec_cuts} }

The final selection criteria have been evaluated in order to
optimize the sensitivity of the experiment. Since the true values
of the oscillation parameters are unknown, we optimized the data
reduction to deliver the most stringent upper limit on \sit\ for a
given \Dm\ under the assumption that there are no \numubnueb\
oscillations. Even a small oscillation signal would then first
materialize as a much less stringent upper limit then the
experimental sensitivity. The optimized cuts were obtained by
simulating and analyzing experimental outcomes with different cuts
leading to different event statistics \cite{Armfzka}. It turned
out that the achievable sensitivity only slightly depends on the
variation of reasonable data cuts.

The final data cuts are as follows: Accounting for the ISIS time
structure, the \pos \ from \numubnueb\ oscillations must be
detected in the time interval of $0.6 <t_{pr.}<10.6 \ \us$ after
beam on target, in which $84.0\%$ of all \numub\ are expected. The
lower time cut of 600 nanoseconds is chosen to eliminate any
contributions from beam correlated fast neutrons (see
section~\ref{subsec:br_neutrons}). The lower cut on the visible
energy deposit $E_{pr.}$ of a positron candidate is 16~MeV. This
energy cut eliminates the neutral current contributions \NC\ to
the neutrino induced random background (fig.~\ref{fig_rnd}) and
also suppresses the soft component of the cosmic induced
background (fig.~\ref{fig_cbred}). No fiducial volume cut for the
\pos \ is applied.

The time difference for the delayed neutron capture event is
restricted to the interval $5<\Delta t < 300 \ \us$. Here, the
lower time cut is fixed by a minimum hardware deadtime after the
electronic read-out of the prompt event. The upper time cut at
$\Delta t< 300 \ \us$ is an outcome of the  MC procedure mentioned
previously and reflects the different time distributions of
delayed events from neutron capture ($\tau\approx120 \ \us$) and
from the background reactions of random coincidences (uniformly
distributed) and charged current coincidences ($\tau = 15.9 \
\mbox{ms}$).

The remaining data cuts for neutron capture events are the
coincidence volume of $V_c= 1.3 \ \mbox{m}^3$ and a maximum energy
of the neutron capture event of $E_{del.}< 8.0\ \mbox{MeV}$. Table
\ref{table_cuts} gives a summary of the applied data cuts and the
corresponding efficiencies $\varepsilon$, resulting in a total
efficiency
\begin{equation}
\varepsilon_{tot}({\nueb})=0.192\pm0.0145
\end{equation}
for an oscillation signal at large \Dm .
\begin{table} [h]
  \centering
  \begin{ruledtabular}
  \begin{tabular}{lcc}
  event & data cut &efficiency $\varepsilon$\\ \hline
  & check on &  \\
  & previous history, &  0.709 \\
  & (see sec. \ref{subsec_req}) & \\
  \raisebox{1.5ex}[-1.5ex]{\pos} & $0.6 \ \us \ < \mbox{t}_{pr.}<10.6 \ \us $ & 0.840\\
  & $16\ \mbox{MeV} < \mbox{E}_{pr.} < 50 \ \mbox{MeV} $ & 0.775 \\ \hline
  & $5\ \us\ < \Delta \mbox{t} <300 \ \us $ & \\
  $(n,\gamma)$ & $ \mbox{E}_{del.} < 8.0 \ \mbox{MeV}$ &  0.416 \\
  & $V_c=1.3 \ \mbox{m}^3$ & \\
  \end{tabular}
  \end{ruledtabular}
  \caption{Final data cuts and efficiencies for the \numubnueb\ search. The
  efficiency for the energy cut corresponds to oscillation parameters $\Dm \ge 100$~\eV .}
  \label{table_cuts}
\end{table}

\subsection{Data reduction \label{subsec_data}}

\begin{figure*}
\begin{center}
\includegraphics{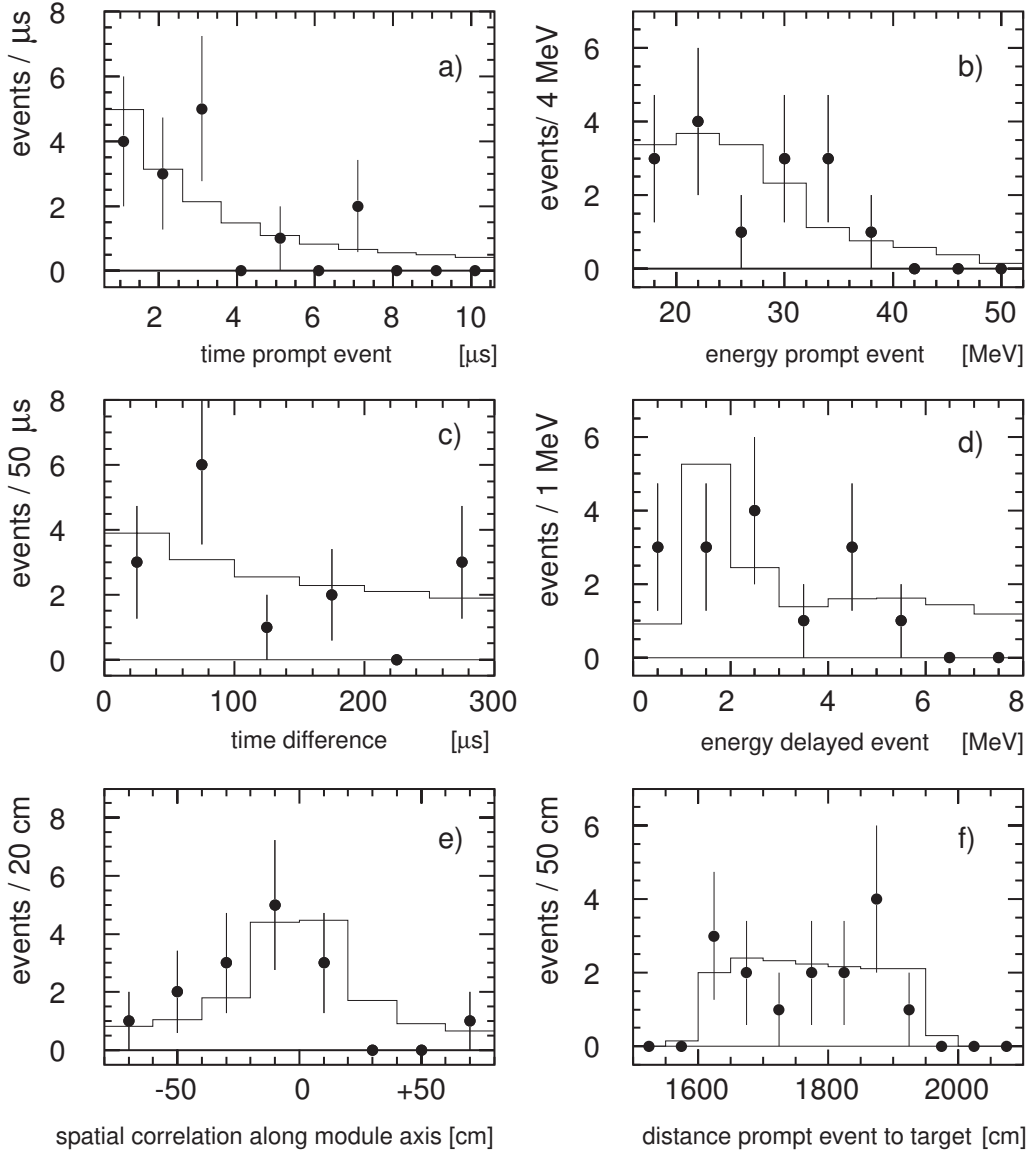}
\caption{Final event ensemble (a) time of prompt events, (b)
energy of prompt events, (c) time difference between prompt and
delayed event, (d) energy of delayed events, (e) spatial
correlation and (f) distance to target of prompt event. The 15
oscillation candidates are in very good agreement with the
background expectation of 15.8 events (solid line).}
\label{fig_signplot}
\end{center}
\end{figure*}

Applying the final selection criteria to the entire KARMEN 2 data
set results in  15 \nueb \ candidate events.\\
The total  background expectation  amounts to $N_{BG}^{exp.}=
(15.8 \pm 0.5)$ events for the components described in section
\ref{sec_bg}. As can be seen from the summarizing table
\ref{bgtable}, the background is dominated by neutrino induced
processes, whereas the cosmic induced background contributes to
only 25\% of the total rate. The relative uncertainty of the
background expectation amounts to 5\%, reflecting the accuracy of
the {\it in situ} measurement of the three dominating background
components in different energy and time windows.
\begin{table*}
  \begin{ruledtabular}
  \begin{tabular}{lcl}
    background & expectation $N_i$ & method of determination\\ \hline
    Cosmic induced background & $3.9 \pm 0.2$ & measured in diff. time window\\
    Charged current coincidences & $5.1 \pm 0.2$ & measured in diff. energy, time windows\\
    \nue \ ind. random coincidences  & $4.8 \pm 0.3$ & measured in diff. time window\\
    \nueb \ contamination            & $2.0 \pm 0.2$ & MC- simulation\\ \hline
    Total background $N_{BG}^{exp.}$ & $15.8 \pm 0.5$& \\
  \end{tabular}
  \end{ruledtabular}
  \caption{Expected background contributions}\label{bgtable}
\end{table*}
Figure~\ref{fig_signplot} shows the spectral distribution of the
15 candidate events with the superimposed background expectation,
normalized to 15.8 events. In each plot the measured data agree
well with the expected background distributions. There are no
obvious deviations from the background expectations, neither for
the prompt nor  delayed events.

Already, the agreement of the number for measured events with the
expected background does not give any  hint for an oscillation
signal within the KARMEN 2 data. In the following, we will set
upper limits on the oscillation parameters, also using spectral
information of the candidate events.
\begin{table*}
   \begin{ruledtabular}
  \begin{tabular}{lcl}
    detection reaction & expectation $N_{sig}$ & neutrino source \\ \hline
    \nuebanp   & $2716 \pm268$ & \numubnueb \ from main target \\
    \nuebanp   &   $73 \pm7$ & \numubnueb \ from $\mu$SR target \\
    \nuebanc   &  $125 \pm17$ & \numubnueb \ from main target \\
    \hline
    Total $N_{sig}^{exp}(\sit=1,\Dm=100\eV)$    & $2913\pm269$ & \\
  \end{tabular}
   \end{ruledtabular}
  \caption{Expected \numubnueb\ oscillation signal for maximal mixing}\label{sigtable}
\end{table*}
\section{Data analysis \label{subsec_ana}}

For \numubnueb\ oscillations with maximal mixing (\sit\,=\,1) and
large mass differences ($\Dm\geq 100 \eV$), an oscillation signal
of ($2913\pm269$) sequences is expected (see table \,
\ref{sigtable}). This number includes a small contribution from
\numub\ produced at the intermediate ISIS $\mu$SR
target\footnote{The $\mu$SR (muon spin resonance) target is
located in the ISIS beam line upstream of the main target at a
distance of 29.2\,m to the KARMEN detector.}. The systematic error
of the oscillation expectation is dominated by the neutrino flux
uncertainty of 6.5\,\% \cite{Bur95} and the error in the
determination of the neutron detection efficiency of 7.0\,\%.

Having measured 15 events with a background expectation of 15.8
events, there is no indication for the presence of an oscillation
signal in the KARMEN data. Ignoring, in a first step, the
spectroscopic information of the measurement and interpreting the
experimental outcome as a pure counting experiment, an oscillation
signal larger than $N_{sig}=7.4$ events is excluded in 90\%
confidence interval (C.I.) \cite{PDG97,Hel83}. However, such a
simplified approach does not make any use of the spectroscopic
quality of the data. In order to extract more information on a
potentially small oscillation signal in the final event ensemble,
a single event based maximum likelihood analysis is applied to
this ensemble.

\subsection{Likelihood analysis}

\begin{figure*}
\begin{center}
\includegraphics{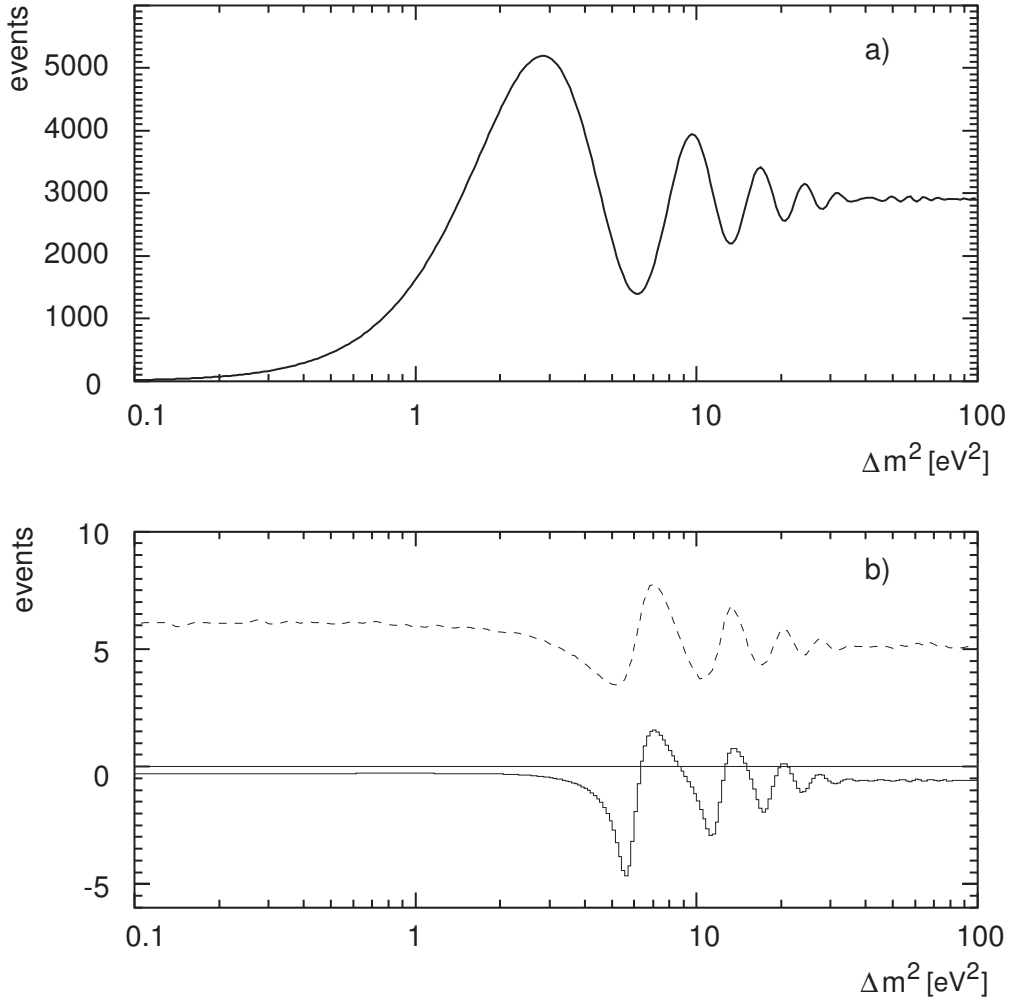}
\caption{(a) Expected oscillation signal for maximal mixing
$\sit=1$. (b) Results of the likelihood analysis: The solid line
shows the best fit of a \numubnueb\ signal in the KARMEN\,2 data.
The dashed line corresponds to the upper bound of the derived 90\%
confidence interval (C.I.) for an oscillation signal. Note that
there is no lower bound of the 90\% C.I. for all \Dm.}
\label{fig_lhresults}
\end{center}
\end{figure*}

The purpose of a maximum likelihood analysis is the separation of
a potential signal from background by maximizing the likelihood
function with regard to some unknown parameters. In this case, the
signal corresponds to (\pos,n) sequences from \numubnueb\
oscillations, the unknown estimators are the oscillation
parameters \sit\ and \Dm . The likelihood function
$\cal{\tilde{L}}$ is defined as:
\begin{equation} \label{lhftil}
{\tilde{\cal L}}(r,\Dm) = \prod^{N_{sample}}_{n=1} \left[r\cdot
f_{sig} (\vec x_n,\Dm) + (1-r)\cdot f_{bg}(\vec x_n)\right]
\end{equation}
with the following definitions:
\begin{itemize}
  \item The event sample with $N_{sample}$=15 candidate events
is characterized by the
  information on the energy $E_{pr}$ and time $T_{pr}$ of the prompt
  event, the energy of the delayed event $E_{del}$ and the time difference $\Delta T$ and
  position correlation $\Delta x$ between prompt and delayed event. This information for each
  event sequence $n$ is represented by the vector $\vec
  x_n=(E_{pr},T_{pr},E_{del},\Delta T,\Delta x)$.
  \item $f_{sig}$ and $f_{bg}$ are the probability density
  functions
  for the vector $\vec x_n$ in case of event $n$ being a signal  or a background event.
  \item The parameter $r$ describes the signal fraction
  in the data and is connected to \sit\ by the linear transformation
  \begin{equation}
    \sit = \frac{r \cdot N_{sample}}{N_{sig}^{exp.}(\sit=1,\Dm)}
  \end{equation}
  with the calculated oscillation signal $N_{sig}^{exp.}$(\sit=1,\Dm) for maximal
  mixing as shown in figure \ref{fig_lhresults} (a).
  \item Assuming no correlation for the $j=5$ observables of $\vec x$, the probability density
  function is factorized to:
  \begin{equation}
  f_{sig}    =     \prod_{j=1}^5 f_{j,sig}
  \end{equation}
  \begin{equation*}
  \qquad \  =  f(E_{pr},\Dm) \cdot f(T_{pr}) \cdot f(E_{del})\cdot f(\Delta T)\cdot f(\Delta
   x)
  \end{equation*}
  \item Due to the  small event sample size of 15
  events, the fit is not performed by varying simultaneously the signal and all background
  components individually. In contrast, the four individual background components are
  added up to one total background component
  \begin{equation}
     f_{bg}=\sum_{i=1}^4 c_i \cdot \left( \prod_{j=1}^5 f_{j,bg_i} \right)
  \end{equation}
  with the coefficients $c_i$ being the expected relative
  contributions of the background channels. The values of $c_i$ are given by the ratio of
  the expected number of background events $N_i$ of each
  component and the total background expectation $N_{BG}^{exp}$ (see table
  \ref{bgtable}): $c_i=N_i/N_{BG}^{exp}$, thereby satisfying the normalization condition
  $\sum_i c_i = 1$.
  \item With the normalization constraint of the probability density function, the parameter
  $r$ determines also the background contribution in the likelihood maximization:
  \begin{equation}
     N_{bg} = (1-r) \cdot N_{sample}
  \end{equation}
\end{itemize}

with the above definitions, maximizing $\cal{\tilde{L}}$ with
regard to \sit\ and \Dm\ is a pure shape analysis and does not
take into account the knowledge of the total background
expectation $N_{BG}^{exp}$. To include this quantitative
information, the likelihood function is weighted with a Poisson
probability term $P_{P}$ computing the probability of measuring
$N_{bg}(r)$ background events for an expectation of $N_{BG}^{exp}$
events:
\begin{equation}
    {\cal L}(r,\Dm) = {\tilde{{\cal L}}}(r,\Dm)  \cdot
    P_{P}\left(N_{bg}(r)|N_{bg}^{exp}\right) 
\end{equation}
with
\begin{equation}
    P_{P}\left(N_{bg}(r)|N_{bg}^{exp}\right) =
    \frac{(N_{bg}^{exp})^{(1-r)N_{sample}}e^{-N_{bg}^{exp}}}{\Gamma(1+(1-r)N_{sample})}
\end{equation}
The expansion in the Poisson probability from the discrete
factorial $n!$ to the Gamma function
$\Gamma(x)=\int_0^{\infty}e^{-t}t^{x-1}dt$ with $\Gamma(n+1)=n!$
ensures a continuous calculation of the Poisson probability for
any signal ratio~$r$.

Maximizing the above defined likelihood function $\cal{L}$ for the
final KARMEN\,2 data results in a best fit for $r$ compatible with
the no-oscillation solution. In fact, the global maximum of
$\cal{L}$ is reached slightly in the unphysical region, at
oscillation parameters:
  \begin{equation} \label{eq_unphys}
  \sit=-2.4\cdot 10^{-3} , \quad \Dm=5.4\,\eV
  \end{equation}
Restricting the analysis to the allowed region, the likelihood
function is found to be maximal at:
  \begin{equation}
  \sit= 8.0\cdot 10^{-4} , \quad \Dm=7.0\,\eV
  \end{equation}
Table \ref{lhvalues} shows the number of (\pos,n) sequences from
\numubnueb\ oscillations of some selected parameter combinations
$k$. Also given are the differences of the likelihood values to
the maximum in the physically allowed region: $-\Delta \ln {\cal
L}=\ln {\cal L}_k-\ln {\cal L}_{max}$. The logarithmic likelihood
value of the best fit differs from the likelihood value for no
oscillation by only 0.21 units.
\begin{table}
  \centering
  \begin{ruledtabular}
\begin{tabular}{cccc}
  $\Dm [\eV]$ & \sit & $N_{sig}$ &
  $-\Delta \ln \cal{L}$ \\ \hline
 \multicolumn{2}{c}{no oscillation} & 0 & 0.21 \\
  0.1 & $-1.7\cdot 10^{-2}$ & $-$0.3 &      0.21 \\
  5.4 & $-2.4\cdot 10^{-3}$ & $-$4.4 &   $-$0.75 \\
  7.0 & $ 8.0\cdot 10^{-4}$ &    1.5 &      0.00 \\
  100 & $ 2.1\cdot 10^{-4}$ & $-$0.6 &      0.18 \\
\end{tabular}
\end{ruledtabular}
 \caption{Signal event numbers for selected oscillation scenarios.
The values of $-\Delta \ln \cal{L}$ indicate the difference of the
likelihood value to the maximum in the physically allowed region
(see text).}\label{lhvalues}
\end{table}
As will be discussed in section~\ref{subsec_comp}, a statistical
analysis of the likelihood function indicates that for boundaries
of 90\% confidence intervals (C.I.), typical differences of
$-\Delta \ln {\cal L}\approx 4-5$ have to be applied. This
underlines the fact that the maximum at
$(\sit,\Dm)_{max}=(8.0\cdot 10^{-4},7.0$\,\eV ) is statistically
in excellent agreement with the null hypothesis of no
oscillations. \\
Furthermore, simulations of comparable event ensembles, with no
oscillation signal but background only, show that a global maximum
at slightly unphysical oscillation parameters as it is in case
here (Eq.\ref{eq_unphys}) is a typical result of the likelihood
analysis of small event samples $\langle N_{sample}\rangle=15.8$.

Figure~\ref{fig_lhresults}(a) shows the expected oscillation event
numbers as a function of \Dm\ for maximal mixing $\sit=1$. In
contrast, figure~\ref{fig_lhresults}(b) demonstrates the results
of the maximum likelihood analysis.
For 90 slices per decade in \Dm , the number of oscillation events
for maximal likelihood $N_{sig}^{max}$ is plotted (solid
histogram). For low as well as high values of \Dm , the
corresponding best fits are almost identical with the physical
boundary, with values of $N_{sig}^{max}=-0.3$ and
$N_{sig}^{max}=-0.6$, respectively. In a region of about $3\le \Dm
\le 30$\,\eV, stronger variations of the energy spectrum of a
potential signal come into play: Since KARMEN has an excellent
energy resolution of $\sigma_E\approx 2$\,\% for positrons with
$E\approx 30$\,MeV, statistical variations in $E_{pr}$ of the
small event sample can be easily interpreted by the likelihood
analysis as modification of the background energy spectrum due to
oscillations with an oscillation length comparable to the distance
target-detector:
\begin{equation} \label{oscil}
 L_{osc} = \frac{2\pi \cdot E}{1.27\cdot \Dm} \approx 17\,{\rm m}
\end{equation}
For energies $12\le E(\numub)\le 52.8$\,MeV, equation
(\ref{oscil}) leads to oscillation parameters of about $3\le \Dm
\le 15$\,\eV\footnote{In more detail, this argument is extended to
the second oscillation mode, with $L_{osc,2}\approx 17/2=8.5$\,m,
which explains the variation up to values of about $\Dm \le
30$\,\eV, as can be seen in figure~\ref{fig_lhresults}.}. It is
important to note that the results given in
figure~\ref{fig_lhresults}(b) for $3\le \Dm \le 30$\,\eV\ are
statistically perfectly compatible with the no-oscillation
solution, as will be discussed in the next section.

\subsection{Upper limits on oscillation parameters \label{subsec_comp}}

Finally, the confidence intervals for the parameters \sit\ and
\Dm\ have to be deduced from the experimental likelihood function.
Recently, there have been discussions \cite{Cer00} about  various
approaches  in order to obtain confidence regions, especially
under the aspects of event samples of low statistics, oscillatory
behavior of the likelihood function as well as parameter
determination near physical boundaries. In the following, we adopt
the Unified Approach \cite{Fel99} which is a frequentist approach
with a specific ordering principle: In the $[\sit,\Dm]$-plane, a
2-dimensional confidence interval (C.I) for the oscillation
parameters is constructed from the comparison of the experimental
likelihood value $\Delta \ln {\mathcal L }=\ln {\cal
L}(\sit,\Dm)-\ln {\cal L}((\sit,\Dm)_{max}$ with the outcome of a
large sample of Monte Carlo simulations of so-called toy
experiments for this term. These simulations are based on the
detailed knowledge of all resolution functions and the spectral
information on the background. In addition, they comprise the
expected experimental signal for an oscillation hypothesis with
given parameters $(\sit,\Dm)$. The hypothesis is then accepted in
the 90\% C.I. if the experimental value does not lie within the
outer 10\% tail  of the simulated $- \Delta \ln {\mathcal
L}$-distribution. For a complete statistical analysis, the entire
parameter space $[\sit,\Dm]$ is scanned to extract the according
region of confidence.

In figure~\ref{fig_lhresults}(b), the result of this approach is
shown in terms of excluded oscillation events. The dashed line
corresponds to the limit of the 90\,\% confidence interval,
excluding larger signal event numbers. For $\Dm =100 \eV$ an
oscillation signal stronger than 5.1 events is excluded in the
90\% C.I., while for low \Dm $<0.1 $ an oscillation signal larger
than 6.0 events is excluded. Though one of the major features of
the Unified Approach is the possibility of extracting lower limits
within the same analysis, no {\it lower} limit of the 90\,\% C.I.
appears, demonstrating the compatibility of the likelihood result
with the no-oscillation hypothesis regardless of the chosen value
for \Dm .

\begin{figure*}
\begin{center}
 \includegraphics{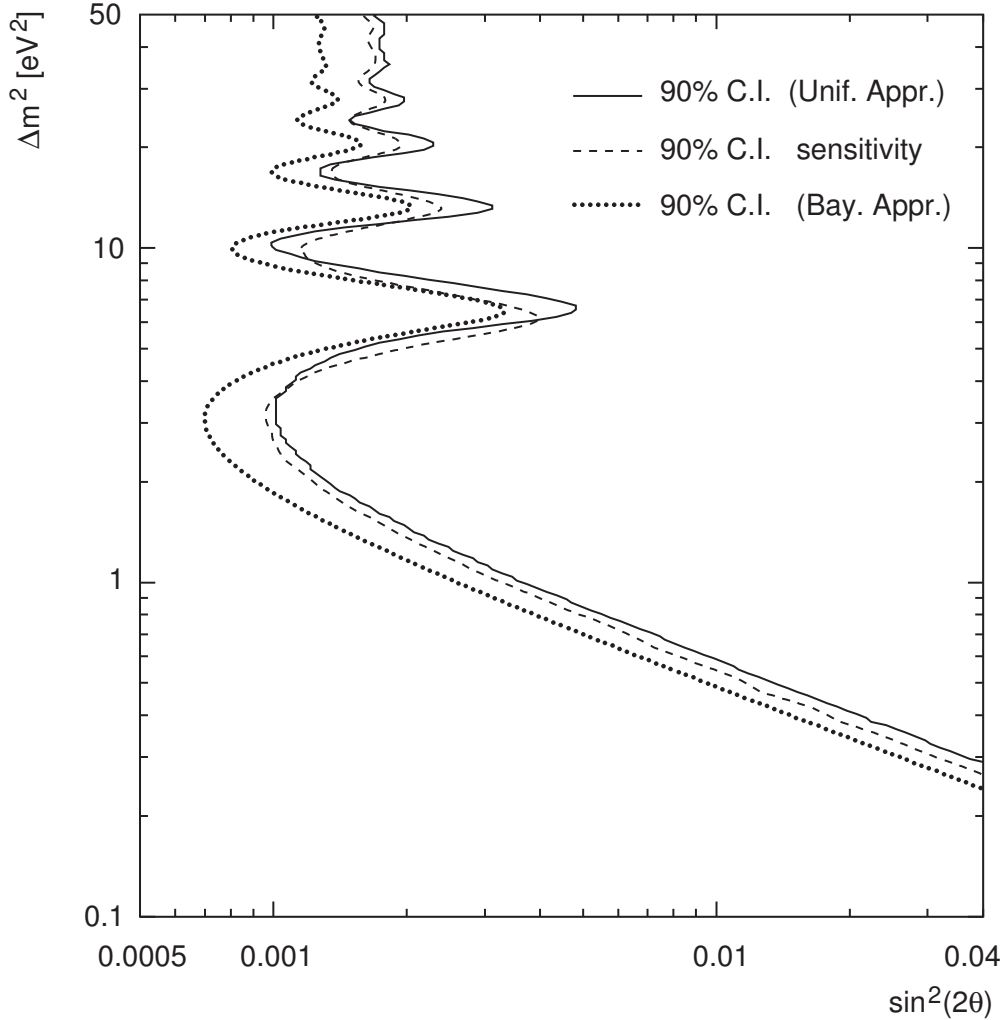}
 \caption{KARMEN\,2 $90\%$ C.I. result deduced with the Unified
  Approach (solid); $90\%$ C.I. sensitivity within the Unified
  Approach (dashed) and $90\%$ C.I. in the Bayesian approach
  (dotted). Regions to the right of the curves are excluded. Note the zoom of the axis in \sit , not reaching up to 1.}
 \label{fig_osziplot1}
\end{center}
\end{figure*}
The exclusion plot in the 2-dimensional $[\sit,\Dm]$-plane
(fig.~\ref{fig_osziplot1})is derived by dividing, for all values
of \Dm , the excluded events (see the solid line in figure
\ref{fig_lhresults} (b)) by the expectation for maximal mixing
(figure \ref{fig_lhresults} (a)). This results in the 90\%~C.I.
limits:
\begin{eqnarray}
 \sit &<& 1.7\cdot 10^{-3} \quad   \
 \Dm \ge 100\,\eV \label{limitresults} \\
 \Dm &<& 0.055\,\eV \quad \quad    \
 \sit = 1
\end{eqnarray}
The entire exclusion curve is shown in figure~\ref{fig_osziplot1}
as solid line, excluding parameter combinations in the area right
to the curve.

An important criterion of an experimental result and a derived
upper limit is the question of how close the limit quoted is to
the experimental sensitivity. Following \cite{Fel99}, the
sensitivity is defined as expectation value for the upper limit of
the 90\,\% confidence interval under the assumption of no
oscillations and is gained by simulations of experiments'
outcomes. The KARMEN\,2 sensitivity as a function of \Dm\ is shown
in figure~\ref{fig_osziplot1} as dashed line. The sensitivity
$\langle{\sit}\rangle$ for $\Dm=100$\,\eV\ amounts to
\begin{equation}
 \langle{\sit}  \rangle = 1.6 \cdot 10^{-3} \quad 90\,\% \ \mbox{C.I.}
\end{equation}

For completeness, we also perform a Bayesian approach to derive an
upper limit on the oscillation parameters \sit\ and \Dm . In the
Bayesian framework, the upper limits for fixed \Dm\ are obtained
by integrating the likelihood function $\cal{L}$. This integration
implies the use of a prior probability density distribution for
\sit\ \cite{PDG01} and decomposes the 2-dimensional problem into a
one-dimensional treatment. We used a uniform prior in a
logarithmic metric of the oscillation parameter \sit . In both the
frequentist and Bayesian approaches, we restrict the parameter
space to the physically allowed region. The Bayesian 90\%~C.I.
approach yields more stringent upper limits shown as dotted line
in figure~\ref{fig_osziplot1} with
\begin{equation}
 \sit < 1.3\cdot 10^{-3} \quad
 \Dm \ge 100\,\eV
\end{equation}

Because of the ambiguities in choosing the probability density
distribution for \sit\ as well as the 2-dimensional oscillatory
behavior of the likelihood function, we do not favor the Bayesian
extraction of confidence intervals but refer to the results
deduced within the frequentist Unified Approach (see
Eq.~\ref{limitresults}). The resemblance of the KARMEN exclusion
curve with its sensitivity curve underlines the fact, that the
likelihood analysis results in no indication of a \numubnueb\
oscillation signal in the KARMEN\,2 data.

\subsection{Comparison with LSND and other experiments \label{subsec:Comp}}

\begin{figure*}[htb]
\begin{center}
\includegraphics{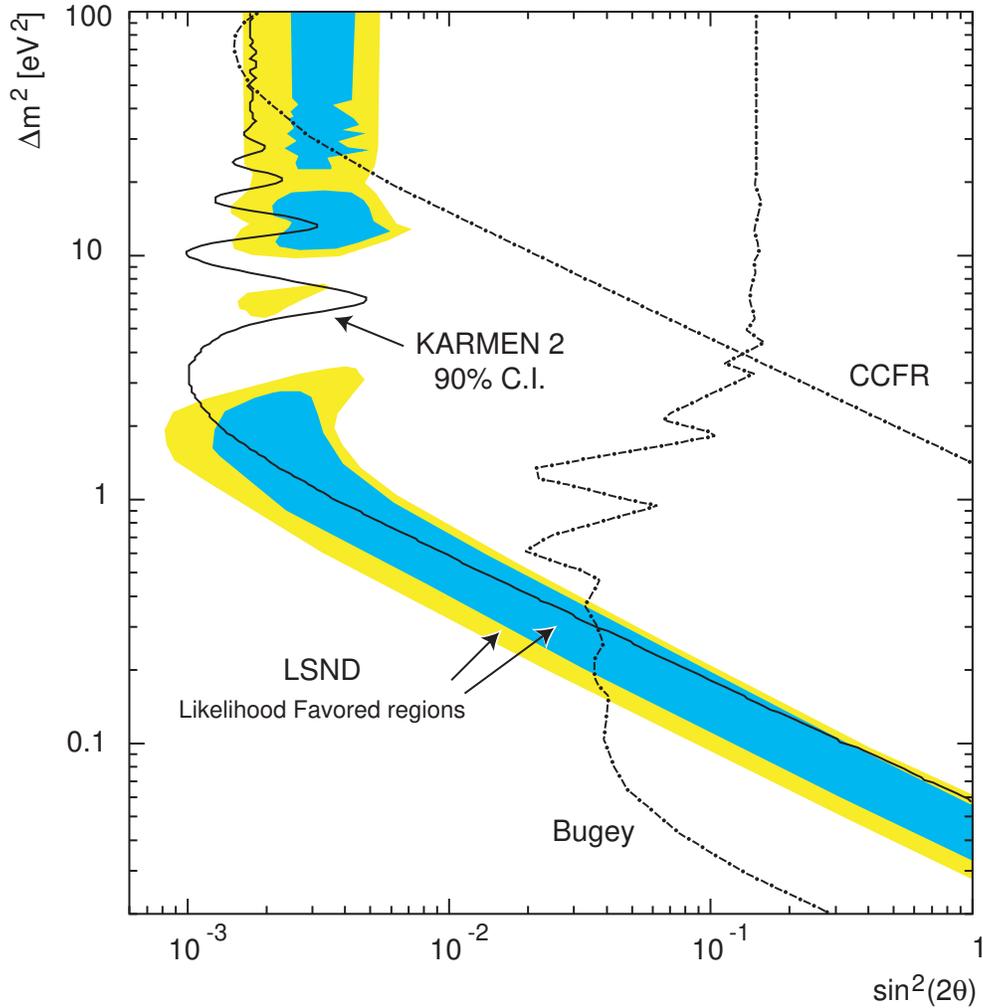}
\caption{Comparison of oscillation searches performed by different
short baseline experiments.} \label{fig_osziplot2}
\end{center}
\end{figure*}

The parameter space for oscillations \numubnueb\ excluded at
90\,\% C.I. by the KARMEN\,2 measurement is shown in
figure~\ref{fig_osziplot2}. The KARMEN result sets the most
sensitive limits so far on \numubnueb\ oscillations in the
parameter region of $0.3 \le \Dm\ \le 30$\,\eV . At higher \Dm\
values, the area right to the right to the exclusion curve is also
excluded by a combined \numunue\ and \numubnueb\ search of CCFR
\cite{ccfr}. The search for \nueb\ disappearance at the Bugey
reactor \cite{bugey} excludes small \Dm\ values, at large
amplitudes $A>0.03$. \footnote{Note that, in a complete 3-- or
4--neutrino mixing scenario, due to the \nueb\ disappearance
search of \protect\cite{bugey} the oscillation amplitude describes
a combination of mixing angles different to that of \numubnueb\
appearance experiments such as KARMEN and LSND (see e.g.
\protect\cite{fogli}).}
The parameter area excluded by KARMEN covers large parts of the
parameter combinations favored by the LSND experiment
\cite{LSNDdet}. The LSND result plotted here shows areas obtained
by cutting the experiment's logarithmic likelihood function at
constant values 2.3 and 4.6 units below the likelihood maximum
\cite{lsndfinal}. For values of $\Dm \le 2$\,\eV , the oscillation
signal expected in KARMEN based on the LSND region ($\ln{\cal
L}_{max}-2.3$) corresponds to a range of 3 to 14 oscillation
events. As shown in figure~\ref{fig_lhresults}, a signal larger
than 6 events is excluded at \NCL . At $\Dm \ge 20$\,\eV , the
expected LSND signal of 7 to 13 oscillation events in KARMEN is in
clear contradiction to the KARMEN upper limit of 5.1 (6.5) events
at 90\,\% C.I. (95\,\% C.I.).

These examples based on expected additional \nueb\ events from
\numubnueb\ demonstrate that at smaller values of \Dm\ there is a
restricted parameter region statistically compatible with both
experimental results. At high \Dm\ values, the LSND solutions are
in clear contradiction with the KARMEN upper limit.
\section{Conclusion}

Results based on the entire KARMEN2 data set collected from 1997
through to 2001 have been presented. The extracted candidate
events for \nueb\ are in excellent agreement with background
expectations showing no signal for \numubnueb\ oscillations. A
detailed likelihood analysis of the data leads to upper limits on
the oscillation parameters \sit\ and \Dm\ excluding parameter
regions not explored analyzed by other experiments.

These limits  exclude large regions of the parameter area favored
by the LSND experiment. A more quantitative statistical statement
on the compatibility between  KARMEN and LSND has to be based on a
combined statistical analysis of both likelihood functions
\cite{eitel}. Such a detailed joint statistical analysis has been
performed \cite{Joint}.

The negative search for \nueb\ from muon decay at rest presented
here sets also stringent limits on other potential processes of
\nueb\ production such as lepton family number violating decays
\mupdeb\ or neutrino oscillations \nuenueb\ which will be
discussed in a separate paper. Future experiments such as the
MiniBooNE experiment at Fermilab \cite{MiniBooNE} aim at
investigating the LSND evidence and the oscillation parameters not
yet excluded by the \numubnueb\ search presented here.

\section{Acknowledgements}

We gratefully acknowledge the financial support from the German
Bundesministerium f\"ur Bildung und Forschung (BMBF), the Particle
Physics and Astronomy Research Council (PPARC) and the Council for
the Central Laboratory of the Research Councils (CCLRC). In
particular, we thank the Rutherford Appleton Laboratory and the
ISIS neutron facility for  hospitality and steady support during
years of data taking.
\newpage

\end{document}